\documentstyle[aps,floats,epsfig,epsf,preprint]{revtex}

\begin{document} 
 
\draft
\title{Giant vortices in the Ginzburg-Landau description of
superconductivity}
\author{Vincent Hakim$^{(1)}$, Ana\"{e}l Lema\^\i tre$^{(2)}$\footnote{
 Present address: Department of Physics, University of California,
 Santa Barbara, CA  93106}
 and Kirone Mallick$^{(3)}$
 }
\address{$^{(1)}$ Laboratoire de Physique Statistique\footnote{
associ{\'e} au CNRS et aux Universit{\'e}s Paris VI et VII},
Ecole Normale Sup{\'e}rieure, 24 rue Lhomond, 75231 Paris Cedex 05,
France  \\
$^{(2)}$Service de Physique de l'\'Etat Condens\'e, CEA Saclay,
 F-91191 Gif-sur-Yvette Cedex, France\\
$^{(3)}$ Service de Physique Th\'eorique, CEA Saclay,
 F-91191 Gif-sur-Yvette Cedex, France 
}
\maketitle 
\begin{abstract}
 Recent experiments on mesoscopic samples 
 and theoretical considerations lead us to analyze multiply
 charged ($n>1$) vortex solutions of the Ginzburg-Landau equations for
 arbitrary values of the Landau-Ginzburg parameter $\kappa$. For
 $n\gg 1$, they have a simple structure and a free energy ${\cal F}\sim n$.
 In order to relate this behaviour  to the classic  Abrikosov
 result ${\cal F}\sim n^2$ when $\kappa\rightarrow +\infty$, we consider
 the limit where both $n\gg 1$ and $\kappa\gg1$, and   obtain
 a scaling function of the variable $\kappa/n$   that   describes  the
 cross-over between these  two  behaviours of 
 ${\cal F}$. It is then   shown  that a
small-$n$ expansion can also be performed and the first two terms of this
expansion are calculated. Finally, large and small $n$ expansions are given for
recently computed phenomenological exponents characterizing the free
energy growth with $\kappa$ of a giant vortex. 
\end{abstract}
 \section{Introduction}

Recent experiments  \cite{geim1,geim2,deoprl1} have
demonstrated that vortices of charge $n>1$, which are unstable in macroscopic
type II superconductors \cite{gussig},
 can exist  in  mesoscopic superconducting samples and can even be
favored 
over 
configurations with multiple singly charged vortices. 
 The
Ginzburg-Landau
description of superconductivity 
\cite{ll,degennes,sarma} appears to be an adequate framework to
analyze
these results. 
The magnetization
response of a mesoscopic sample can be  analytically understood
by  adding  a surface energy contribution
to the Ginzburg-Landau free  energy of a giant-vortex 
in an infinite system  \cite{akmal1,akmal2}.
Moreover, full numerical solutions of the Ginzburg-Landau equations
accurately reproduce the experimental findings \cite{sp1,sp2,peetreview}.

Motivated by these
experimental and theoretical results, we analyze in the present
paper
`giant'(i.e. $n>1$) vortex solutions of the  Ginzburg-Landau equations.
Well-known 
analytical results  have been obtained in the London limit
when the Ginzburg-Landau parameter $\kappa\to \infty$
\cite{abrikosov,sarma}
and at the special dual-point value $\kappa = \frac{1}{\sqrt{2}}$
\cite{sarma,bogo}. Here,
we take advantage of the supplementary  parameter $n$, the vortex
charge, and provide a simple
analysis of the giant vortices for arbitrary values of $\kappa$.
 We begin  in section 2, by  recalling
 the  Ginzburg-Landau equations and some
elementary properties of their solutions.
In section 3, we consider the large vorticity
limit
$n\gg 1$. The vortex structure takes the form of a circular
normal core separated by a sharp boundary from the outside superconducting
medium. As a consequence, 
the  free energy $\cal{F}$ of a giant
vortex is  found to be 
 proportional to its charge $n$ in contrast to the classic
Abrikosov's result ${\cal F}\sim n^2$ in the London limit.
In order to relate the two results, we consider in section 4,
the double limit in which both the vorticity
$n$ and the  Ginzburg-Landau parameter $\kappa$ are large and
we obtain the scaling form ${\cal F}/2\pi\sim n\kappa \Phi(\kappa/n)$.
The function $\Phi$ provides an explicit interpolation between
Abrikosov's result for $\kappa\gg n$ and
the result of section 3 which is valid in the opposite  limit $n\gg\kappa$.
To complete our analysis of giant vortex solutions, we consider
in section 5,  the   somewhat more formal  $n \to 0$
limit, which
  has the advantage to be amenable to a simple and  systematic expansion
 scheme;  the free energy is obtained
 up to order $n^2$.
In section 6, we compare our perturbative results (valid for 
 arbitrary values of $\kappa$)
to the known exact results at the dual point $\kappa=1/{\sqrt{2}}$.
Besides checking consistency, we obtain a  perturbative
expansion of the free energy in the vicinity of the dual point and the
large-$n$ expansion of phenomenological 
exponents  introduced  previously   and  computed 
numerically \cite{akmal2}.
The large $n$ expansion proves to be
fairly accurate for $n$ values as small
 as 3 or 4.

\section{The Ginzburg-Landau model  of superconductivity}

In the Ginzburg-Landau description of  superconductivity
 \cite{ll,degennes,sarma}   the two unknown fields are the 
 complex order parameter  $\psi = |\psi| e^{i\chi}$ 
 and the potential vector ${\vec A}$   with 
 ${\vec{\nabla}}\times{\vec A} = {\vec B}$, where ${\vec B}$ is 
 the local magnetic induction. These fields satisfy 
 the following equations:
\begin{eqnarray}  
  -  ({\vec{\nabla
}} - i\frac{2\pi}{\phi_0}{\vec A})^2 \psi & = & 
 \frac{1}{\xi^2} \psi 
(1 -  |\psi{|^2}) ,  \\ 
    {\vec \nabla} \times {\vec \nabla} \times {\vec A}
 & = & \frac {|\psi|^2}{\lambda^2} ({\phi_0}\frac{{\vec{\nabla}}\chi}
{2\pi} - {\vec A} ).
\label{eqcomplete}
\end{eqnarray} 
 Here, the flux quantum $\phi_0$ is given by 
  $\phi_0 = {hc}/{2e}, $  and 
 the two characteristic lengths $\lambda$ (penetration depth or
London length)  and $\xi$ 
 (coherence length)  appear as phenomenological parameters. 
 The Ginzburg-Landau   parameter $\kappa$ is defined as their ratio,
   $\kappa = \frac{ \lambda} {\xi}$.
 We shall    measure  lengths in units of 
 $\lambda { \sqrt 2} $, the magnetic field 
 in units of $ H_c/{\sqrt{2}\kappa} = {\phi_0}/{4\pi\lambda^2}$
  and the vector potential 
 in units of  ${\phi_0}/{2\sqrt{2}\pi\lambda}$. 
 The Ginzburg-Landau  free energy in  units of ${H_c^2}{\xi^2}/{4\pi}$   
 is then given by:
\begin{equation} 
{\cal F} =  {\int} {1 \over 2} B^2 + {\kappa^2} 
 (1 - |\psi|^2 )^2 + 
|({\vec{\nabla}} - i{\vec A})\psi {|^2}    . 
\label{energ} 
\end{equation}
A giant vortex of vorticity $n$ is a   solution of the
Ginzburg-Landau equations with cylindrical symmetry
$ \psi(r, \theta)   =   f(r)\exp( in\theta) , $ 
where $(r,\theta)$ are polar coordinates in the plane. The potential vector $
\vec A$ can be chosen to lie in the plane and to have only  a 
non-zero angular component $A_{\theta}(r)$.
It is convenient to introduce an  auxiliary function $g$ that represents
  the difference between the flux through  a disk of radius $r$
 and  the total flux,
 \begin{equation} 
    g(r)     =    r A_{\theta}(r)  - n  \,\,\,\, {\it i.e.}  \,\,\,\,
      B = \frac{1}{r}  \frac{ d g}{dr} .  \label{defg}
 \end{equation}
 The free energy can be rewritten in terms of $f$ and $g$ as
 \begin{equation}
 \frac{1}{2\pi} {\cal F} = \int_{0}^{\infty} \left\{
  \left(\frac{df}{dr}\right)^2 + \frac{f^2g^2}{r^2} +
  \kappa^2 (1 - f^2)^2 + \frac{1}{2r^2}\left(\frac{dg}{dr}\right)^2
  \right\} rdr .
  \label{free2}
  \end{equation}
  The allied Ginzburg-Landau equations reduce to 
\begin{eqnarray} 
  \frac{ d^2 f}{dr^2}  + \frac{1}{r}  \frac{df}{dr} 
 - \frac{1}{r^2} fg^2  &=&  - 2 \kappa^2 f (1 - f^2), \label{gl1}  \\ 
    \frac{ d^2 g}{dr^2} -  \frac{1}{r}  \frac{ d g}{dr}  &=& 2 f^2 g .
  \label{gl2}
\end{eqnarray} 
It is useful to note that a very simple form of the free energy \cite{sarma}
 is obtained
by an efficient use of (\ref{gl1}) and (\ref{gl2}),
\begin{equation}
  \frac{1}{2\pi} {\cal F} =  2  \kappa^2 \int_{0}^{\infty}  r dr (1 - f^2) .
  \label{identiteB} 
 \end{equation} 
 
A vortex of charge $n$ corresponds to functions  $f$ and $g$ which satisfy
$f(0)=0$  and $g(0)=-n$ at the origin and which obey
$f(\infty) = 1$  and $g(\infty) = 0$ at infinity.
Linearization of Eqs.~(\ref{gl1}, \ref{gl2}) around $f=1$ and $g=0$ shows
that there exists two exponentially growing and two exponentially decaying
spatial modes at $r=\infty$. In the same way, linearization for $r$ close to
zero shows that there is one diverging mode with $f\sim r^{-n}$ and one neutral
mode corresponding to changes in the vortex charge (note that at the level
of Eqs.~(\ref{gl1}, \ref{gl2}), $n$ appears simply  as a parameter and is
not constrained to be  an integer). Once one requires 
the diverging mode at $r=0$ to be
absent and $g(0) = -n$, the expansion of $f$ and $g$  around $r=0$ depends
on two arbitrary constants
 \begin{eqnarray} 
       f(r)  &=&  \left(\frac{r}{{\cal R}}\right)^n  + {\cal O}(r^{n+1}), 
 \label{localf} \\
       g(r)  &=& -n\left( 1 -  \left(\frac{r}{r_0}\right)^2\right) 
+ {\cal O}(r^{2n+2}).
  \label{localg}
  \end{eqnarray}
 The length scales  $r_0$ and 
 ${\cal R}$  are   uniquely determined by the cancellation of the two
divergent modes at $r=\infty$\cite{note1}; they 
 cannot be calculated  from a local analysis near 0. Their determination
requires  the behaviors around $r=0$ and $r=\infty$
 to be connected. This can be
done  numerically for arbitrary  parameter values or analytically 
when $n$ is either large or small as shown in the following
sections.

One can note from the definition of $g$ (Eq.~(\ref{defg})) 
that  $r_0$  is simply related to the
 value  of the magnetic field  at the position of the vortex:
 \begin{equation}
    B(0) = \left( \frac{1}{r} \frac{dg}{dr} \right)_{r=0} = \frac{2n}{r_0^2}.
  \label{champorig}
\end{equation}

 \section{A giant vortex in the large vorticity limit}

 We first consider the structure of a giant vortex of charge $n$ for
$n\gg 1$ and begin with simple estimates. It seems intuitively clear
that the vortex core grows with its charge. So, for $n\gg 1$, one
expects the vortex core to be much larger than the London 
 penetration length. As a consequence,
the magnetic induction should be approximately constant over the core.
 For a vortex of charge $n$ and core size $r_c$,
 one expects therefore $|B|=2n/r_c^2$ in the vortex core (the total flux divided
by the core area with the chosen normalization) and $B=0$ outside.
The magnetic energy of such a configuration is approximately $\int B^2/2\simeq
2\pi n^2/r_c^2$.
Correlatively, one expects $\psi=0$ in the vortex core and $\psi=1$ outside
and the corresponding energy contribution $\int \kappa^2 (1-|\psi|^2)^2\simeq
2\pi \kappa^2 r_c^2$.
The total Ginzburg-Landau energy (\ref{energ}) of such a configuration can
therefore be estimated to be 
\begin{equation}
\frac{1}{2 \pi}{\cal F}\simeq \frac{n^2}{r_c^2} +\frac{\kappa^2}{2} r_c^2 +
\sigma(\kappa) r_c ,
\label{simpest}
\end{equation}
where the last term on the right-hand side
 has been added to take into account the 
interfacial energy of the transition layer between the vortex core and the
superconducting bulk. Minimizing Eq.~(\ref{simpest}) with respect to $r_c$
gives the dominant order estimates for the vortex size 
$r_c$ and for its free energy ${\cal F}$
 \begin{eqnarray}
r_c^2 &\sim&\frac{\sqrt{2} n}{\kappa} ,
\label{valr0}
\\
 \frac{1}{2\pi} {\cal F} 
 &\simeq&  n \kappa\sqrt{2} + \left( \frac{\sqrt{2}n}{\kappa}\right)^{1/2}
 \sigma(\kappa) +{\cal O}(1) .
\label{approxA}
  \end{eqnarray}
The magnetic field inside the vortex
core has the constant value 
\begin{equation}
B=\sqrt{2} \kappa [1+\frac{\sigma(\kappa)}{2\kappa^2 r_c}] .
\label{gtcond}
\end{equation} 
That is, in the vortex core
$B$ is equal to $H_c$  plus  a Gibbs-Thomson correction
as expected at a curved normal/superconducting boundary 
(see e.g. \cite{dorsey2,chapman}).
In the next subsections, we first present numerical solutions of the 
Ginzburg-Landau equations for various values of  $n$ which confirm this
simple picture of the giant vortex. This picture
 is  then derived from a direct analysis
of the Ginzburg-Landau equations in the large $n$ limit.

 \subsection{Numerical results}

Numerical solution of
Eqs.~(\ref{gl1},\ref{gl2}) consists in  solving 
 a two-point boundary value
problem (at $r=0$ and $r=\infty$)
and can be achieved using several methods \cite{numrec}.

Shooting is generally easy to implement  and robust
when  one mode needs to
be cancelled at a boundary: a free parameter is adjusted (for example by
dichotomy) at the other
boundary until the required cancellation is obtained. In the present
case, two diverging modes need to be cancelled at infinity which is
more difficult to achieve. We kept a simple one parameter shooting by
integrating
from a large $r=r_{\infty}$ toward $r=0$. Of the two free parameters
at $r_{\infty}$ one was used to cancel the diverging mode at $r=0$. The
value of $g$ at $0$ was then given as a function of the other parameter (which
could be adjusted to obtain a particular value of $n$ when desired).
 
We  also implemented  a relaxation method. 
The system of ordinary differential  equations is  replaced by  a 
 set of  finite difference equations satisfied on a mesh of points,
 and the boundary conditions just appear as equations satisfied by the points
 located at the extremities of the mesh.
 A  multi-dimensional version of  Newton's
 method provides a solution of these finite difference equations
 by an iterative procedure.  This method
 requires the inversion of a  matrix  of  size   proportional
 to the number of points in the mesh.  Because  this matrix is block diagonal,
 it  can be inverted in a very efficient manner. However,  the efficiency
 of the relaxation method  depends strongly  on the starting point
 of the iteration: for example, it is  useful
 to  take the profile of a $n$-vortex as an initial guess for 
 the $(n+1)$-vortex solution of the Ginzburg-Landau equations.

Numerical solutions for different values of $n$ are shown in Fig.~1. The plot
of $f$ shows a well-defined
 interface between a normal region (where $f=0$)  and a superconducting region
 $(f=1)$ (see Fig.~1a). The interface width 
 does not increase with $n$ but remains  of the order of $1/\kappa$. 
As expected, the interface
 has a well-defined limiting shape:
 in Fig.~1b, 
 the curves  represented in Fig.~1a   are  shifted
  in such a way that they   take the value
 $1/2$ at the point $r = 0$.  
The  shifted  curves superpose on  one another almost perfectly.
  Only the curve for the small
 value $n =4$ shows a noticeable deviation from the limiting shape.

\begin{figure}
\begin{center}
\includegraphics[width=7.cm,height=7.cm,angle=-90]{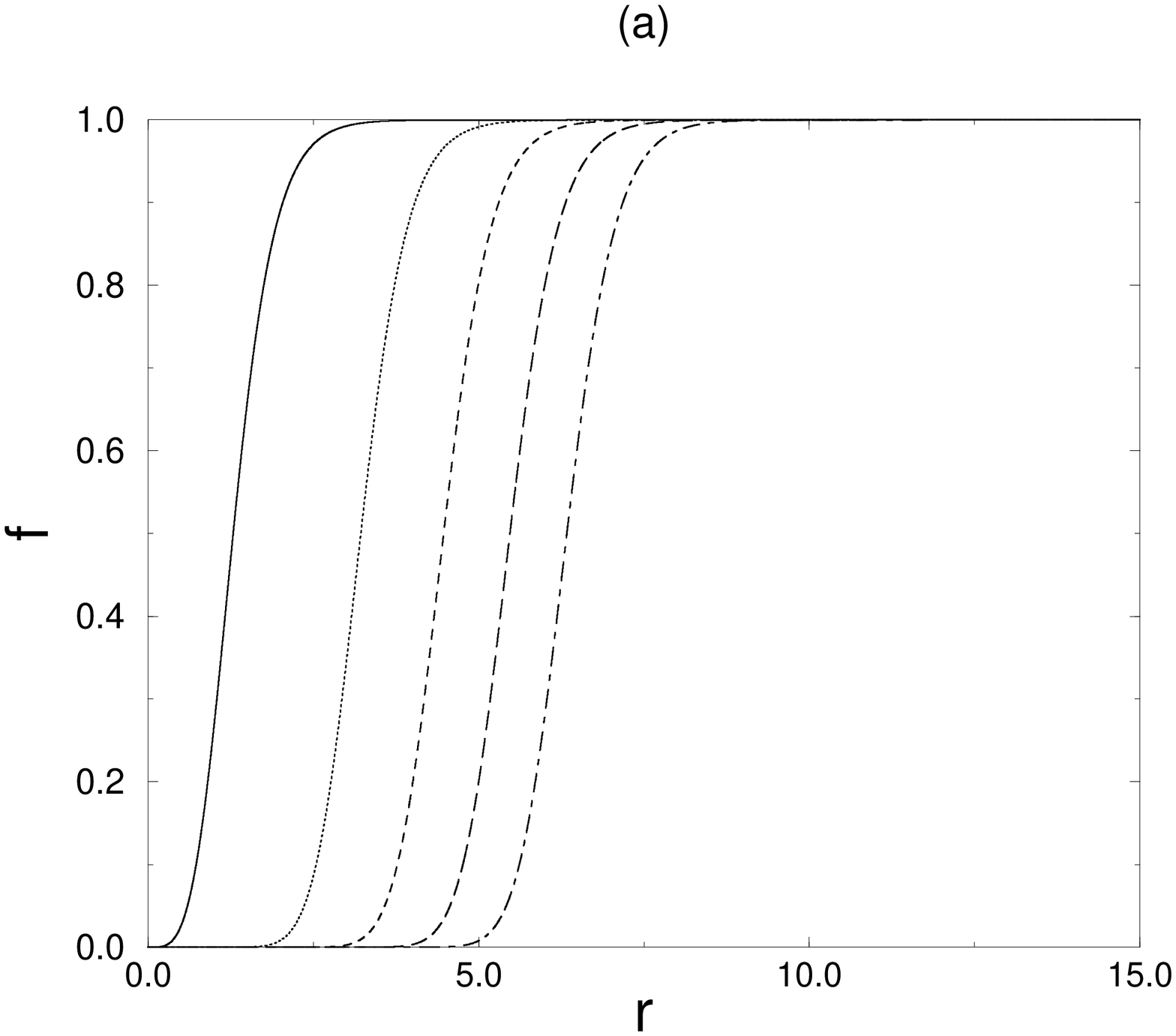}
\includegraphics[width=7.cm, height=7.cm,angle=-90]{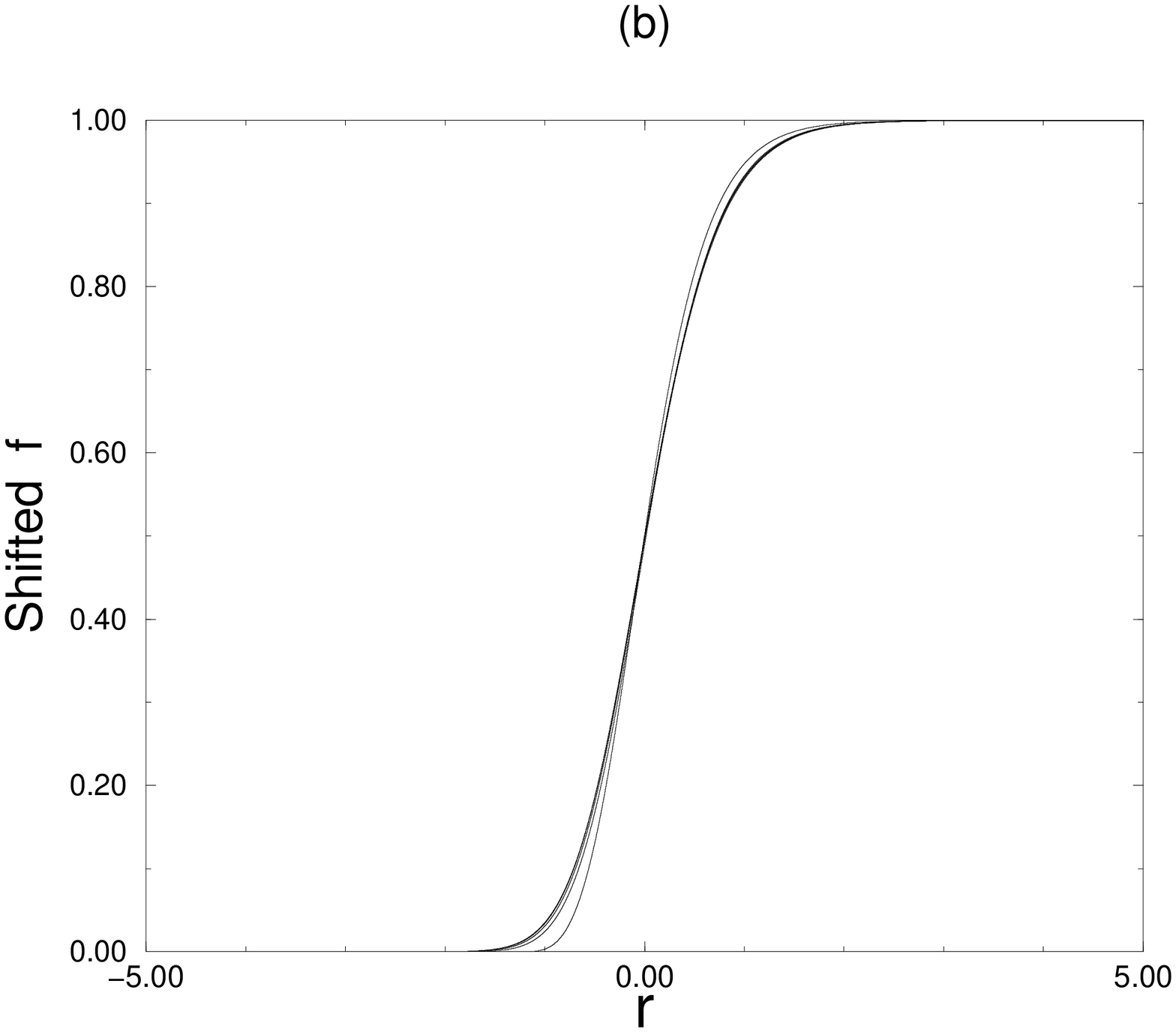}
\end{center} 
\caption{a) The function $f$ for $\kappa$ = 1.5 and  n =
4,16,28,40,52. b)
The different front profiles 
 superposed  on  one another ($f=1/2$ is shifted to  $r=0$).
}
\end{figure}

\begin{figure}
\begin{center}
{\psfig{figure= 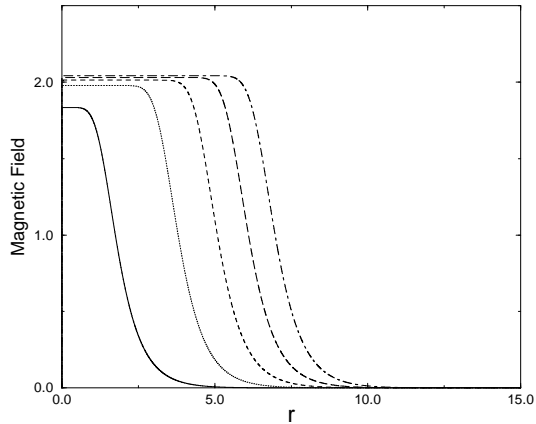,height=7cm,angle=-90}}
\end{center} 
\caption{The magnetic field  for $\kappa^2 = 1.5 $ and $n
 = 4,16,28,40,52.$}
\label{bfig}
\end{figure}

  The  simple normal/superconducting interface picture for $n\gg 1$ is also
confirmed by the graphs of $g$. In Fig.~\ref{bfig},  $ B(r)=1/r\ dg/dr$
is  plotted for different values of $n$. The magnetic induction is
close to $H_c=\sqrt{2}\kappa$ in the vortex interior and quickly decreases to
zero in the interface region. Again,  in the large $n$ limit, the curves
   can be put over  one  another by  the same  shifts  as those used
  to superpose the  different $f$'s in Fig.~1b.

 \subsection{Asymptotic analysis of the Ginzburg-Landau equations for $n\gg 1$}

It is instructive to see how  the large-$n$ giant
vortex structure 
arises in a direct analysis of the Ginzburg-Landau equations
(\ref{gl1},\ref{gl2}).

\subsubsection{The vortex core}
Since $g=-n$ at $r=0$ one expects  $g$ to be of order $n$ in a whole
region near $r=0$. Therefore, at dominant order in $n$, Eq.~(\ref{gl1})
for $f$ reduces to
\begin{equation} 
  \frac{ d^2 f}{dr^2}  - \frac{1}{r^2} fg^2=0 .
\label{fcore}
\end{equation}
The solution of Eq.~(\ref{fcore}) can be obtained in a WKB manner but
the important point is that $f$ is exponentially close to zero in the
whole core region. Eq.~(\ref{gl2}) for $g$ thus simplifies to
\begin{equation}
\frac{ d^2 g}{dr^2} -  \frac{1}{r}  \frac{ d g}{dr}=0 .
\end{equation}
At dominant order in $n$, $g$ is therefore given by
\begin{equation}
g(r)  = -n\left( 1 -  \left(\frac{r}{r_0}\right)^2\right) .
\label{gcore}
\end{equation}
 Although Eq.~(\ref{gcore}) is identical
to Eq.~(\ref{localg}), it should be noted that
 it is not an expansion near zero but the full
function $g$ in the vortex core at dominant order in $n$. Of course,
Eq.~(\ref{gcore}) is nothing else but  the constancy of $B$ in
the vortex core.

Outside  the vortex core the medium is in the superconducting phase,
$f=1$ and $g=0$. In order to match these different solutions of
Eqs.(\ref{gl1},\ref{gl2}), we  analyze the interface region
(i.e. the  boundary-layer in the usual matched asymptotics terminology).

It should be noted that $r_c$ the vortex core size and $r_0$
which is obtained from the second derivative of $g$ at the origin 
(Eq.~(\ref{})) coincide at leading order in $n$. Their difference
is of the order of the planar normal/superconducting interface width 
and depends on the
precise criterion used
to measure the vortex core size (i.e. on the precise definition of $r_c$)
as shown below.

 \subsubsection{Local equations near the interface}
The solution (\ref{gcore}) vanishes for $r=r_0$ and therefore  the assumption
$g\sim n$ under which it was derived breaks down at the interface.
 A different simplification
of the Ginzburg-Landau equations can be made in the vicinity of $r_0$.
 We define  a  new variable $x$, centered around
 $r_0$,  such that 
\begin{equation}
  r = x + r_0 .  \label{shiftvar}
\end{equation}
 We  assume (and will verify at  the end) that  $ r_0 \gg |x| $. Substituting
 (\ref{shiftvar})  into (\ref{gl1},\ref{gl2}) 
 we obtain
 \begin{eqnarray} 
  \frac{ d^2 f}{dx^2} 
+ \frac{1}{r_0}  \frac{df}{dx} 
 - \frac{g^2}{r_0^2} f  &=& - 2 \kappa^2 f (1 - f^2) , 
   \label{recenterf} \\ 
    \frac{ d^2 g}{dx^2}
 -  \frac{1}{r_0}  \frac{ d g}{dx} 
 &=& 2 f^2 g .
      \label{recenterg}
\end{eqnarray}
Comparing the magnitude of the different terms in these simplified 
equations, we obtain
 $f\sim 1, g\sim r_0, x\sim 1$ and find  that the first-order
terms $df/dx$, $dg/dx$ on the l.h.s.~ of (\ref{recenterf},\ref{recenterg})
are subdominant. Introducing the  rescaled functions
 $f_0 = f$ and  $G_0=g/r_0$, one
derives   at dominant order in $n$ (indicated by
the subscript $0$) 
the `inner' equations
 which describe  the  behaviour of the order parameter and the
 vector-potential in the interface region 
\begin{eqnarray} 
  \frac{ d^2 f_0}{dx^2}    &=& 
- 2 \kappa^2 f_0 (1 - f_0^2) + G_0 ^2 f_0 ,  \label{inngl1}  \\ 
    \frac{ d^2 G_0 }{dx^2}   &=& 2 f_0^2 G_0 .
  \label{inngl2}
\end{eqnarray}
As expected,
these equations are   identical to the Ginzburg-Landau
equations in one  dimension ({\it i.e.} on an infinite line). Matching
with
the superconducting phase for $x\gg 1$ and the vortex core for $x\ll
-1$ imposes the boundary conditions $f_0(+\infty)=1, G_0(+\infty)=0$
and $f_0(-\infty)=0$.
 This  last condition implies, using (\ref{inngl2}),
that $G_0$  behaves linearly when  $x \to-\infty$
\begin{equation}
G_0 \simeq ax + b .
\label{asymptgam}
\label{localg2}
\end{equation}
In contrast to Eq.~(\ref{gl1},\ref{gl2}), the local system (\ref{inngl1},
\ref{inngl2}) is invariant by translation. Once this is fixed (for example
by imposing the arbitrary criterion $f_0(0)=1/2$) the functions $f_0$ and $G_0$ and 
the constant $b$ are uniquely determined
 (the value of $a$ is independent of this
arbitrary choice).
 Finding  $a$ and $b$ appears
to require  an explicit solution. This can be avoided, at least for
$a,$ because the
reduced system (\ref{inngl1},\ref{inngl2}) 
retains the variational structure of the original
equations and is invariant by translation in $x$. 
The locally conserved quantity $\cal E$
associated with  this continuous symmetry reads
(that is, the magnetic
field in the vortex core)\cite{ll}
\begin{equation}
{\cal E} =  \frac{1}{2} \left(\frac{df_0}{dx} \right)^2 + 
    \frac{1}{4} \left(\frac{d G_0}{dx} \right)^2 - 
   \frac{1}{2} G_0^2f_0^2 - \frac{\kappa^2}{2} (1 - f_0^2)^2 .
\label{hamiltonien}
\end{equation}
Conservation of $\cal E$ between $-\infty$ and $+\infty$
leads to:
 \begin{equation}
   0 = {\cal E}(+\infty) = {\cal E}(-\infty) = a^2/4 - \kappa^2/2  .
\end{equation}
Thus,  $a =\sqrt{2}\kappa$ and  $B=H_c$ in the
normal phase at dominant order in $n$, as expected.
\subsubsection{Matching the vortex core and the interface region}
In order to determine the constant $r_0$ and the interface position,
we have  to match the solutions in the vortex core and in the
interface
region.

 Rewriting (\ref{localg}) in terms of the local variable $x$ leads to 
\begin{equation}
 g \simeq    -n \left( 1 - ( 1 + \frac{x}{r_0})^2 \right)
 \simeq   \frac{2n}{r_0} x     \,\,\,\,   \hbox { hence  }  \,\,\,\,
      G_0 = \frac{g}{r_0}  \simeq   \frac{2n}{r_0^2} x .
 \label{localg3}
 \end{equation}

The family of solutions of (\ref{inngl1},\ref{inngl2}) is $f_0(x-x_c),
G_0(x-x_c)$ with the asymptotic behavior  
\begin{equation}
G_0(x-x_c)=\sqrt{2} \kappa (x-x_c)+b
 \label{localg4}
 \end{equation}
 Identifying the two  local expressions (\ref{localg3}) and (\ref{localg4})
 for the   function
  $g$ in their common region of validity
 $r = r_0 + x $ with $ |x|  \ll r_0$, and $x \to -\infty$ determines
$r_0$ and $x_c$
\begin{eqnarray} 
r_0^2 &\simeq&\sqrt{2}n/\kappa \\
x_c &=&\frac{b}{\sqrt{2} \kappa}
\label{xc}
\end{eqnarray}

\begin{figure}
\begin{center}
\includegraphics[width=7.cm,height=7.cm,angle=-90]{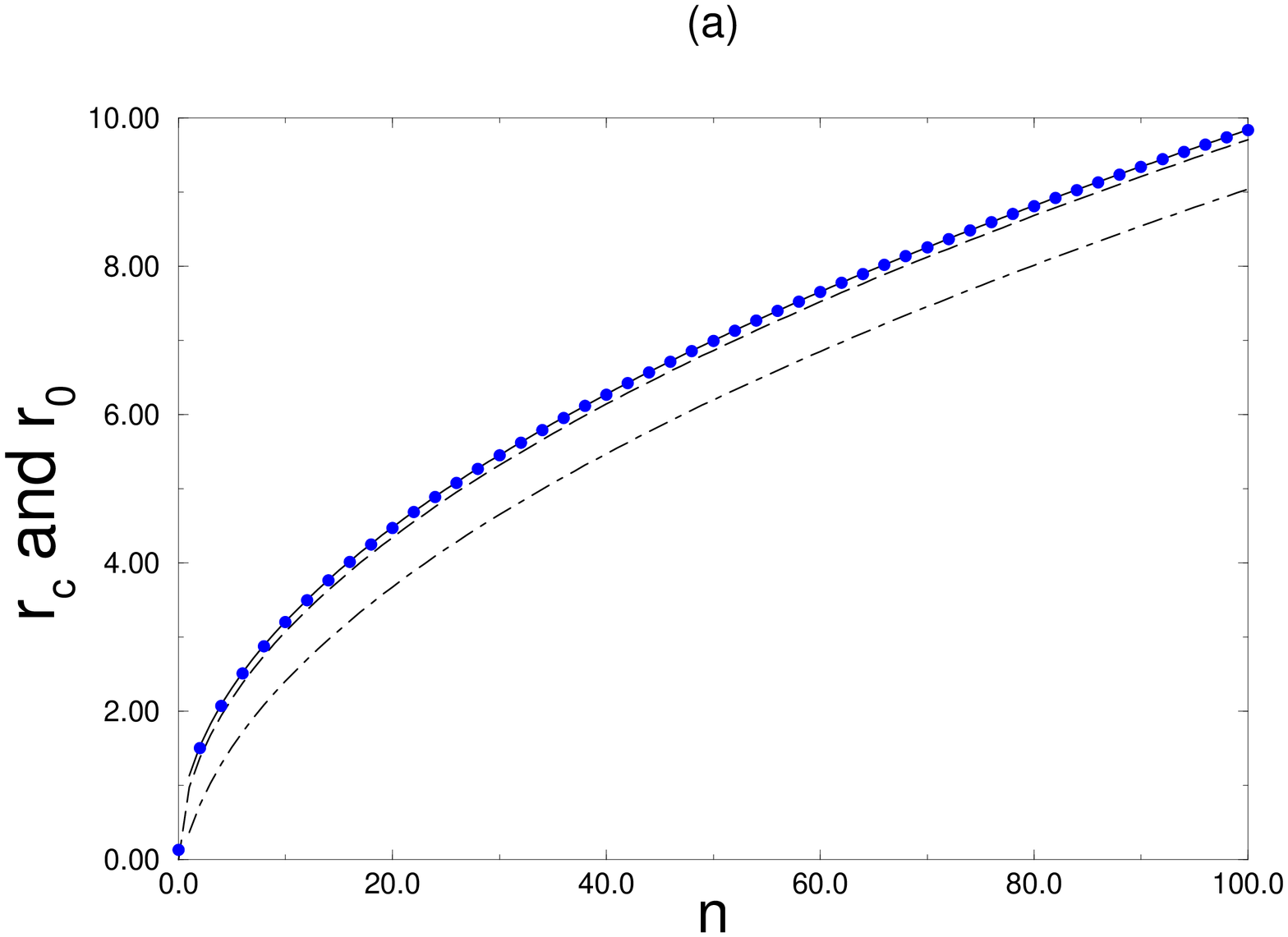}
\includegraphics[width=7.cm, height=7.cm,angle=-90]{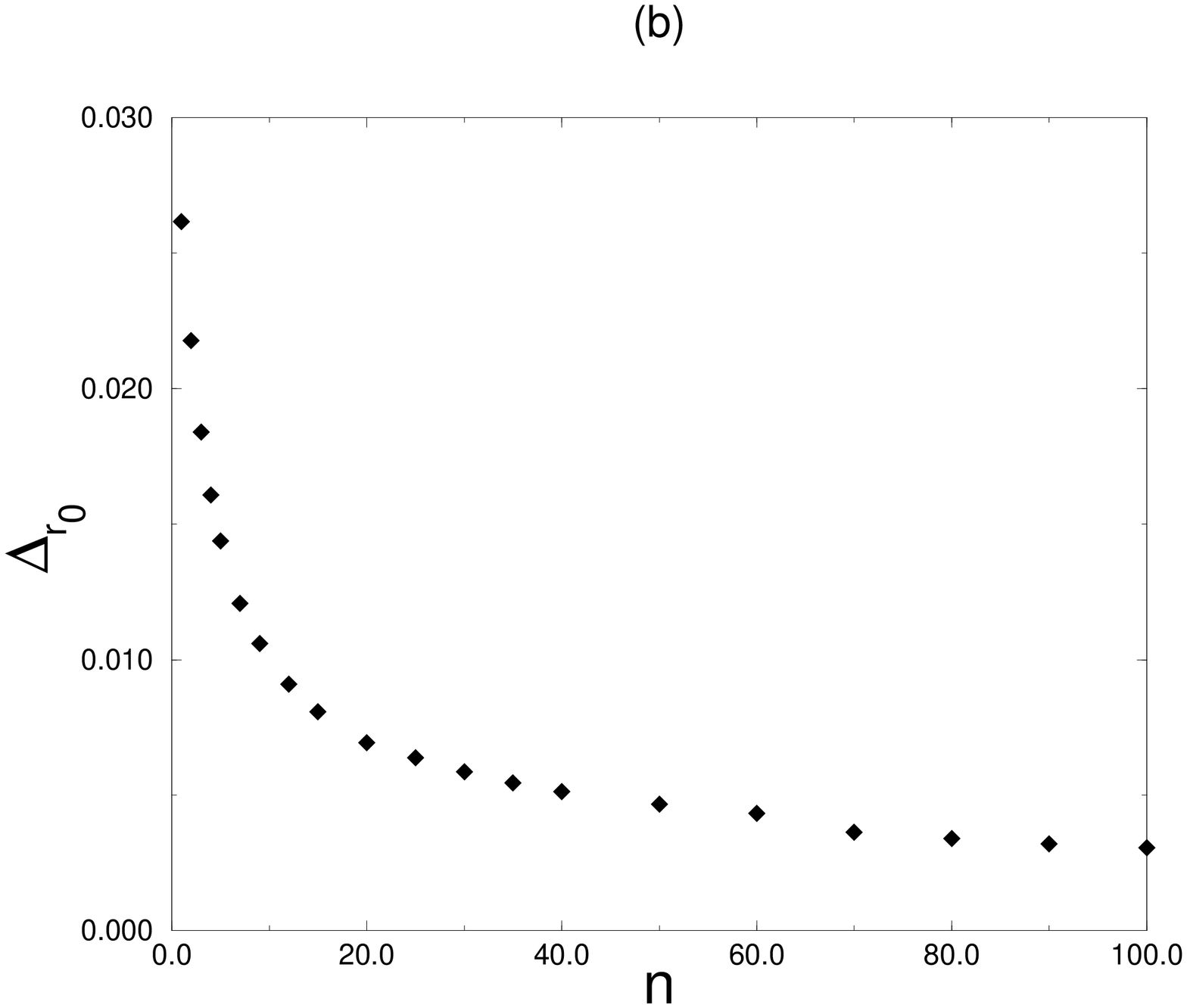}
\end{center} 
\caption{a) The vortex-core size  $r_c$ (dot-dashed curve) and the
 length  $r_0$ (solid line curve)   are  plotted for $\kappa = 1.5$
 and  for  values  of $n$ ranging from 1 to 100.  The dashed curve is the 
 large-$n$ asymptotic  estimate (\ref{valr0}). The dots 
 represent  the next-to-leading order  asymptotic expression
 (\ref{r0ordre1}), using the numerical  value $\sigma(1.5) \simeq -1.172 . $
  b) Difference $\Delta_{r_0}$  between  $r_0$ and
  its  asymptotic expression (\ref{r0ordre1}).  }
\label{frontpos}
\end{figure}

The large-$n$ asymptotic  estimate of $r_0$ (which agrees with 
Eq.~(\ref{valr0})) is
compared with
numerical results in Fig.~\ref{frontpos} a.  

\subsubsection{Large $n$ limit of the energy}
\begin{figure}
\begin{center}
{\psfig{figure= 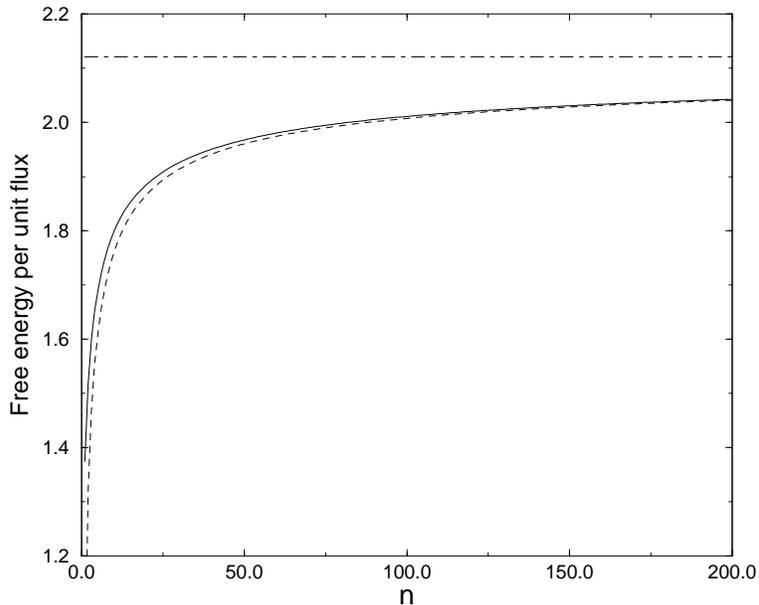,height=10cm,angle=-90}}
\end{center} 
\caption{ Free energy  $\frac{\cal F}{2\pi n}$ as
 a function of $n$ for  $\kappa$ = 1.5
(solid  line). The  horizontal dot-dashed line is   the  asymptotic value
  $\kappa\sqrt{2} \simeq 2.121$. The dashed  curve
 is  the large $n$ behaviour  (\ref{approxA}), using the  
  numerical value $\sigma(1.5) \simeq -1.172 . $}
\label{figf}
\end{figure}

The  free energy of a giant vortex can now be calculated  for $n\gg 1$.
 Using (\ref{gl1})  and integrating by parts, the free energy (\ref{free2}) 
 becomes:
 \begin{equation} 
  \frac{1}{2\pi} {\cal F} =  \int_{0}^{\infty}  r dr 
  \left\{  \frac{B^2}{2}   +  \kappa^2 - \kappa^2 f^4    \right\}  .
 \end{equation}
Completing the square  with  the first two terms
 inside the brackets  and using
 relation (\ref{defg}) between $B$ and $g$, leads to
\begin{equation}
 \frac{1}{2\pi} {\cal F} = n \kappa\sqrt{2} +  \int_{0}^{\infty}  r dr
\left\{ \frac{1}{2} \left( \frac{1}{r}\frac{dg}{dr} -  \kappa\sqrt{2} \right)^2  -  \kappa^2 f^4  \right\} .
\label{identiteA}
 \end{equation}
 The function 
\begin{equation}
 S(r) =  \frac{1}{2} 
 \left( \frac{1}{r}\frac{dg}{dr} -  \kappa\sqrt{2} \right)^2  -  \kappa^2 f^4 ,
\label{denstension} 
 \end{equation}
  is non-zero only in the vicinity of $r_0$.
Hence,  $S$ can be considered 
  to be  a function of $x = r - r_0$.
 Replacing  the functions $f(r)$ and $g(r)$ in (\ref{denstension})
 by their dominant order approximations
 $f_0(x-x_c)$ and $G_0(x-x_c),$  respectively,  solutions of
 (\ref{inngl1}, \ref{inngl2}), we  obtain
\begin{equation}
\int_{0}^{\infty}  r S(r) dr 
 \simeq    r_0  \int_{-\infty}^{\infty} S(x) dx  =
    r_0  \int_{-\infty}^{\infty} \left\{   \frac{1}{2}
 \left(\frac{d G_0}{dx} -  \kappa\sqrt{2} \right)^2  -  \kappa^2 f_0^4 
  \right\} dx    =  r_0 \sigma(\kappa) .
 \label{energsurface}
\end{equation}
 The last integral  is nothing else  but the surface energy  $\sigma(\kappa)$
  of a one-dimensional  domain  wall between normal and
 superconducting phases \cite{ll,fetter,dorsey1} (see appendix \ref{stapp}).
 Using   (\ref{energsurface}) in (\ref{identiteA})  gives exactly
 the large $n$ expression (\ref{approxA})
 of the free energy of a $n$ giant-vortex with $\sigma(\kappa)$ the
surface energy of a one dimensional normal/superconducting interface.
  The expression (\ref{approxA}) is valid
 for any value of $\kappa$ in the limit
 $n \to \infty$. It is compared to the numerical results (for
$\kappa= 1.5$ ) in  Fig.~\ref{figf}.

 \subsubsection{Finite-$n$ correction to $H_c$}
The leading finite-$n$ correction
to $B$ in the vortex core can be derived from the  previous results.
 Integrating   (\ref{identiteB})  by parts leads to
 \begin{equation} 
 \frac{1}{2\pi} {\cal F} =   \kappa^2 \int_{0}^{\infty}  r^2 \phi(r) dr 
  \,\,\,  \hbox{ with } \,\,\,  \phi(r) =  2 f(r) \frac{df}{dr}  .
\label{identiteB2}
\end{equation}
 The function $\phi(r)$ is localized at the position of the front
 and its width is of the order $1 /\kappa$. Using the local variable
 $ x = r -r_0$ defined in (\ref{shiftvar}), the equation (\ref{identiteB2})
 can be rewritten, in the large $n$ limit,  as:
 \begin{equation} 
 \frac{1}{2\pi} {\cal F} =   \kappa^2 r_0^2 + 2 \kappa^2 r_0
 \int_{-\infty}^{\infty} x  \phi(x) dx   + {\cal O}(1) .
 \label{identiteB3}
\end{equation}
The integral in Eq.~(\ref{identiteB3}) can be expressed  in terms
of the interface energy (see appendix \ref{stapp}):
  \begin{equation} 
  2 \kappa^2 \int_{-\infty}^{\infty}\phi(x) x dx =  2 \kappa^2
 \int_{-\infty}^{\infty} \left( (1 - f_0^2) - \frac{1}{\kappa\sqrt{2}}
 \frac{d G_0} {d x}  \right) dx  = \frac{3}{2}  \sigma(\kappa) .
\label{formulesimple}
   \end{equation} 
  Comparing (\ref{identiteB3})  with (\ref{approxA}) provides
  the large $n$ expansion  of $r_0$  up to the second order  
  \begin{equation} 
  r_0 = \left( \frac{ \sqrt{2}n}{\kappa}\right)^{1/2}  -\frac{ \sigma(\kappa)}
  {4\kappa^2} + {\cal O}\left( \frac{1}{n^{1/2}} \right)  .
  \label{r0ordre1} 
    \end{equation}
The vortex size $r_c$, defined by the criterion $f(r_c) = 1/2$,
   can be meaningfully calculated  to
 the same order:
 \begin{equation} 
 r_c=r_0+x_c
 \label{vortsize}
  \end{equation}
with $x_c$ given by Eq.~(\ref{xc}). 

  In order to verify numerically these relations, we
    first  computed  the value of the magnetic
  field  at the origin using  the following relation,
  that derives from   (\ref{defg}) and (\ref{gl2}):
 $$ B(0) =  - 2 \int_0^\infty \frac{f^2 g}{r} dr .$$
  This allowed us to obtain   $r_0$, from (\ref{champorig}),  
 with a good precision.
   The vortex size $r_c$ was  found   from the equation 
 $f(r_c) = 1/2$. In Fig. 3a,  the two numerical curves for
  $r_0$ and $r_c$  are plotted as a function of $n$.
 To the leading order in $n$, these  two
 curves differ by a constant $x_c \simeq -0.797\ldots$,    function
 of $\kappa$ only,  as predicted by  Eq. (\ref{vortsize}).
 Solving  independently  the `inner'  Ginzburg-Landau
equations  (\ref{inngl1},\ref{inngl2}), we found  
 that the  unique value of $b$ such  that $f_0(0)= 1/2$  is
 given by $b \simeq 1.691\ldots $ Hence  the relation (\ref{xc})
 between $x_c$  and $b$ is verified.
 We  also tested  the accuracy  of the
 large $n$ asymptotic expansion of $r_0$ to the first order (\ref{valr0})
 and to the second order (\ref{r0ordre1}). The difference between
$r_0$   and  Eq.  (\ref{r0ordre1}),  hardly
 visible in Fig.~3a,  is plotted in Fig.~3b.  The agreement  is very good.

 Substituting this value of $r_0$ in  (\ref{champorig}), we obtain
the field value inside the core 
 \begin{equation} 
 B(0) =   \kappa\sqrt{2}+ \frac{\sigma(\kappa)}{2^{3/4} \sqrt{\kappa n}}
  +  {\cal O}\left( \frac{1}{n} \right)  .
 \label{formuleB0}
\end{equation} 
 Hence, when $n \to \infty$, the field at the origin is indeed
 the  thermodynamic critical field $H_c=\kappa\sqrt{2}$ with a
 Gibbs-Thomson correction as expected (\ref{gtcond}).

  \section{Relation to Abrikosov's formula}

\begin{figure}
\begin{center}
{\psfig{figure= 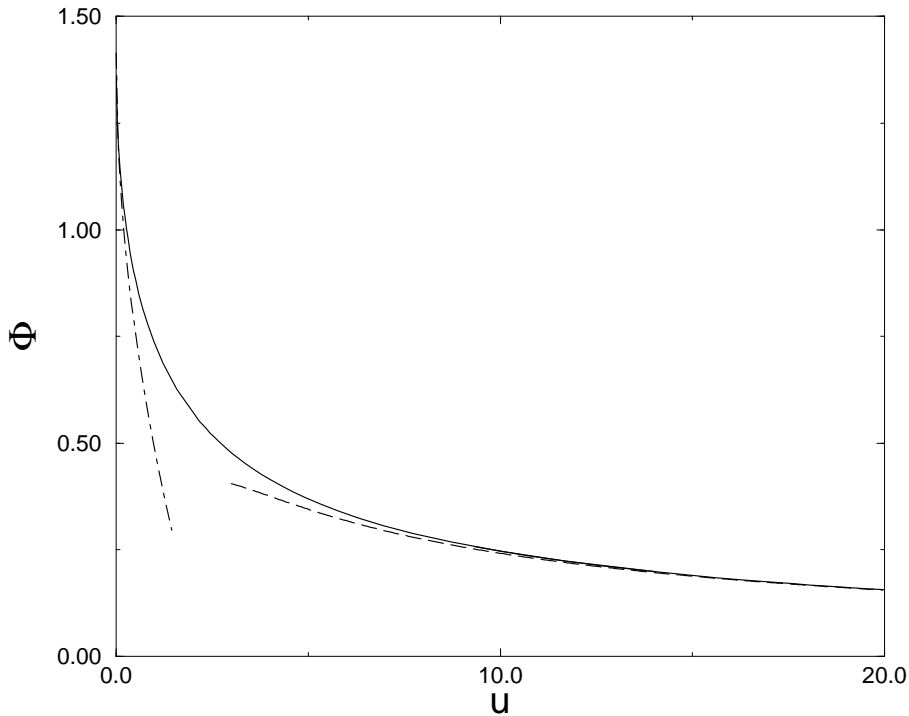,height=12cm,angle=-90}}
\end{center} 
\caption{The scaling function $\Phi$ as a function of the variable $u$
 (solid line). The dot-dashed curve  and the dashed curve represent,
  respectively,   the  large  $u$  (\ref{numul1})
  and small $u$ (\ref{Enupetit}) approximations. }
\label{figphi}
\end{figure}

 The self-energy of a unit  vortex was 
 first  calculated by Abrikosov \cite{abrikosov,sarma} in the 
 large $\kappa$ limit
\begin{equation}
 \frac{1}{2\pi} {\cal F} =   \log \kappa + 0.081 .
\label{Abriko}
 \end{equation}
      In the previous section we have studied the case 
 $n \to \infty$  keeping  $\kappa$   fixed and finite.  In order to 
 relate  our results  to the classical calculation
 of Abrikosov, we  consider in this section the  double limit
 $n \to \infty$ and $\kappa \to \infty$   keeping   the ratio 
 $u = \kappa/n$  finite and fixed. With these assumptions, we
 shall see that the Ginzburg-Landau equations decouple and that a scaling
 form is  obtained for the free energy
\begin{equation}
\frac{1}{2\pi} {\cal F}= n \kappa \Phi(\kappa/n) .
\label{phidef}
\end{equation}
The function $\Phi$ is plotted in Fig.~\ref{figphi}.

\subsection{The double  limit $n\gg 1$ and $\kappa \gg 1$}
\begin{figure}
\begin{center}
{\psfig{figure= 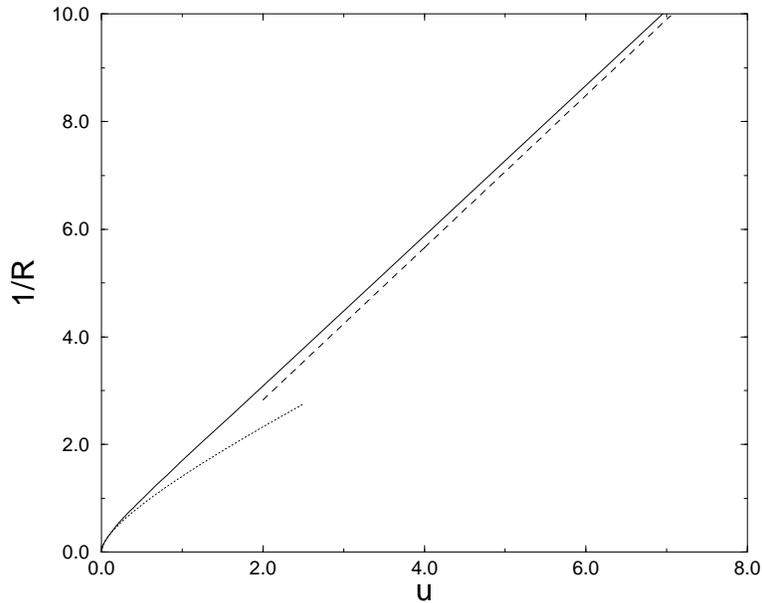,height=10cm,angle=-90}}
\end{center} 
\caption{The inverse of $R$ plotted as a function of the scaling variable $u$ 
  (solid line). The dashed curve represents
 the linear behaviour of $1/R$ for large values of $u$  (\ref{Rgdu}),
 and the dotted curve shows the asymptotic behaviour
 for small values of $u$   (\ref{Rupetit}). }
\label{figR}
\end{figure}

 After rescaling the function $g$ by a factor  $1/n$,
 and introducing the ratio  $u = \kappa/n$, 
 the Ginzburg-Landau equations become:
  \begin{eqnarray} 
 \frac{1}{n^2} \left(\frac{d^2 f}{dr^2} + \frac{1}{r} \frac{df}{dr} \right)
  &=&  - 2 u^2 f (1- \frac{1}{2 u^2 r^2} g^2 - f^2), \label{abgl1}  \\ 
    \frac{ d^2 g}{dr^2} -  \frac{1}{r}  \frac{ d g}{dr}  &=& 2 f^2 g ;
  \label{abgl2}
\end{eqnarray}
the boundary conditions are  $f(0) = 0$  and $g(0) = -1$ at the origin, 
and  $f(\infty) = 1$  and $g(\infty) = 0$ at infinity.
In the $n \to \infty$ limit, 
there are two domains where $f$ is slowly varying  and the derivative
terms in Eq.~(\ref{abgl1})
can be neglected 
 \begin{eqnarray} 
f & = & 0 \,\,\,  \hbox{ for }\,\,\,   r \le R , \nonumber\\  
         f  & = &  \left( 1 - \frac{g^2}{2u^2r^2} \right)^{1/2} \,\,\,
 \hbox{ for }\,\,\,    r \ge R .
 \label{eqlimf}
 \end{eqnarray}
 As  shown below,   these
two domains  match through a boundary layer at $r=R$ where $f$ has
a rapid variation but remains small. Matching the two slowly
varying expressions (\ref{eqlimf})
of $f$  requires that
 $f(R) = 0$ in the second one and gives
 \begin{equation}
  g(R) =  - \sqrt{2} u R  . 
   \label{gcont1}
  \end{equation}
Substituting  (\ref{eqlimf}) in (\ref{abgl2})   
 leads to  the following closed  equations  for  $g$:
  \begin{eqnarray}
\hbox{ for }\,\,\,     r \le R \,\,\,\, \,\,\,\,
  \frac{ d^2 g}{dr^2} -  \frac{1}{r}  \frac{ d g}{dr}   & = & 0  , 
 \label{eqlimg1} \\ 
  \hbox{ for } \,\,\,  r \ge R  \,\,\,\, \,\,\,\,
   \frac{ d^2 g}{dr^2} -  \frac{1}{r}  \frac{ d g}{dr}   & = & 
   2 g  \left( 1 - \frac{g^2}{2u^2r^2} \right)  .
\label{eqlimg2}
 \end{eqnarray}
  From (\ref{eqlimg1}), one has 
  \begin{equation}
  \hbox{ for }\,\,\,     r \le R \,\,\,\, \,\,\,\,  g(r) =  - 1 +  \left(\frac{r}{r_0}\right)^2  .
  \label{parabg}
 \end{equation}
Continuity of
  the function $g$ and of  
 its derivative  at $R$, and equation  (\ref{gcont1}) imply that
 \begin{eqnarray}
    g(R) = -1 +   \left(\frac{R}{r_0}\right)^2 = - \sqrt{2} u R ,
   \label{gcont} \\
    g'(R) = 2 \frac{R}{r_0^2} = \frac{2}{R} - 2 \sqrt{2} u , 
\label{dgcont}
 \end{eqnarray}
(the second equality of (\ref{dgcont})  is derived using (\ref{gcont})).
 The  scaling limit  for the free energy
 when  $n \to \infty$ and $\kappa \to \infty   $ is obtained 
 from (\ref{identiteB}) and (\ref{eqlimf})
 \begin{equation}
  \frac{1}{2\pi} {\cal F} =   n \kappa \left( u  R^2 +
     \frac{1}{u}  \int_{R}^{\infty} \frac{g^2}{r} dr  \right) 
  =    n \kappa u \left(  R^2 + 2 \int_{R}^{\infty} r h^2(r)  \right) ,
    \label{scalenerg}
  \end{equation}
 where  we have introduced a   function  $h$ :
 \begin{equation}
      g  \equiv  h { \sqrt{2} u r} .
 \label{defh}
  \end{equation}
This provides the explicit expression of the interpolating function
$\Phi$ (\ref{phidef})
 \begin{equation}
\Phi(u)= u \left(  R^2 + 2 \int_{R}^{\infty} r h^2(r)  \right) .
\label{phiexpl}
\end{equation}
In order to   calculate explicitly $\Phi$
 one has  to determine the position $R$ of the front and  the function
$h$ as a function of $u=\kappa/n$.
 In  terms of   $h$, Eq.~(\ref{eqlimg2})  becomes 
 \begin{equation}
  \frac{ d^2 h}{dr^2} + \frac{1}{r} \frac{ d h}{dr}   -\frac{1}{r^2} h
   = 2 h ( 1 - h^2) ,
\label{eqlimh}
\end{equation}
 and,  from (\ref{gcont1}), the   boundary conditions 
 for $h$ are
 \begin{equation}
        h(R) = -1  \,\,\,\,\,  \hbox{ and } \,\,\,\,\,
      h(\infty) = 0 .
   \label{bcondh}
 \end{equation}
 Relation  (\ref{dgcont}) implies, however,  one more  condition
 on $h$:
 \begin{equation}
     h'(R) = \frac{\sqrt{2}}{u R^2} - \frac{1}{R} \,\,  .
 \label{bcsupplem}
\end{equation}
 The differential equation (\ref{eqlimh}) with the two  boundary conditions
 (\ref{bcondh}) and the supplementary condition (\ref{bcsupplem})
 is overdetermined: there is  a unique value of $R$
 such that  all the  conditions can be  satisfied. Indeed,   
  equation    (\ref{eqlimh})  with    the boundary
   conditions   (\ref{bcondh}) has  a unique solution
   for any given value of $R$, {\it i.e.}  for a given $R$,
 the value of $h'(R)$ is unique
 and  can easily be computed numerically by solving  (\ref{eqlimh}) 
   by  a shooting method;
 once  $h'(R)$ is known, 
  $u$ is determined  as  a function of $R$  from (\ref{bcsupplem}) 
\begin{equation}
   \frac{u}{\sqrt{2}} = \frac{1}{ R + R^2 h'(R)} .
  \label{inverser}
 \end{equation}
 Inverting this relation  gives  $R$ as a function of $u$
 (Fig.~\ref{figR}) and the function $\Phi$ using (\ref{phiexpl}).
 This calculation   can be performed
 analytically when $u$ is  either very large  or very  small as  shown below.

Before   considering  these limits, we complete the above
analysis by giving the boundary-layer equation satisfied by $f$ in the
neighbourhood of $r=R$. It is convenient to introduce the local
variable
$x=r-R$. Assuming that $f$ remains small in the neighbourhood of
$r=R$ and keeping  the dominant contribution of each term of
 Eq.~(\ref{abgl1}), we obtain  
\begin{equation}
 \frac{1}{n^2} \left(\frac{d^2 f}{dx^2} + \frac{1}{R} \frac{df}{dx} \right)
  =  2 u^2 f (-2 h'(R) x + f^2).
 \label{innab} 
\end{equation}
A consistent dominant balance between the different terms of Eq.~(\ref{innab})
is obtained when $f/(x^2 n^2)\sim f x\sim f^3$. Therefore, as assumed,
$f$  evolves on a short scale $x\sim n^{-2/3}$ around $r=R$ but 
 remains small $f\sim n^{-1/3}$. With the rescaled coordinate  and
function,
$\xi= x (4u^2 h'(R))^{1/3} n^{2/3}, F=(n u)^{1/3}(\sqrt{2} h'(R))^{-1/3} f$,
the boundary-layer equation reads 
\begin{equation}
\frac{d^2 F}{d\xi^2}=-\xi F +F^3  .
\label{innab1}
\end{equation}
The two different functions of Eq.~(\ref{eqlimf})
can be matched through the solution of (the Painlev\'e) Eq.~ (\ref{innab1}). 

\subsection{The large $u $ limit $(\kappa\gg n\gg 1)$ }

   Recalling  that $R$ and $r_0$ are positive,
  we  see from  (\ref{dgcont}) that $  R \le {1}/({\sqrt{2} u}) . $
Therefore, when $u$ tends  to infinity, $R$ tends to 0. In the large
 $u$ limit, equations  (\ref{eqlimg1}) and (\ref{eqlimg2})  reduce  to:
  \begin{equation}
\frac{ d^2 g}{dr^2} -  \frac{1}{r}  \frac{ d g}{dr}  = 2 g ,
  \label{grandu}
\end{equation}
 with the boundary conditions $ g(0) = -1 $  and $g \to 0$
 at infinity. The solution of this equation is readily obtained:
 \begin{equation}
  g(r) = - r\sqrt{2} K_1(r\sqrt{2})  \,\,\,\, \hbox{ and therefore } \,\,\,\,
    h(r) = - K_1(r\sqrt{2})/u .
  \label{solgdu}
 \end{equation}
 where $K_1$ is the modified Bessel function of order 1.
 Using  the behaviour of  $K_1$ in the vicinity of zero 
 (see Appendix B), we deduce that  
 \begin{equation}
 h'(R) \simeq    \frac{ 1}{\sqrt{2} u R^2}  .
 \label{hprimegdu}
  \end{equation}
 Substituting (\ref{hprimegdu}) 
  in (\ref{inverser})  leads to
 \begin{equation}
   R \simeq \frac{1}{\sqrt{2} u} .
\label{Rgdu}
 \end{equation}
  Inserting (\ref{solgdu}) and (\ref{Rgdu}) in the 
 the expression (\ref{scalenerg}) for the free energy,
  we obtain
\begin{eqnarray}
  \frac{1}{2\pi} {\cal F}  &\simeq&  \frac{n \kappa}{ u}
    \left[\frac{1}{2}+\int_{1/ u }^{\infty} r ( K_1(r))^2 dr \right]= 
    \frac{n \kappa }{2u }\left\{1+ \frac{1}{u^2}
\left[ K_0 \left( {1\over u}\right)
        K_2\left( {1\over u}\right)     -   K_1\left( {1\over u}\right)^2
     \right]\right\}  ,  \nonumber \\
&\simeq&    \frac{n \kappa}{ u}
 ( \ln u + \ln 2 - \gamma ) \simeq  n^2 
  \left( \ln\left(\frac{\kappa}{ n} \right) + 0.116 \right) , 
\label{numul1}
 \end{eqnarray}
where the $K_{\nu}$ are modified Bessel functions and $\gamma$ is 
 the  Euler constant \cite{abram}. This relation
 generalizes the classical result (\ref{Abriko}) of Abrikosov
 to the case of a giant vortex. 
  Indeed, the structure of the giant  vortex is very
similar to the $n=1$ case with $f$ varying on a fast scale of order
$n/\kappa$ and $g$ on a slow scale of order one. However, the
structure of $f$  is 
 simpler  in the  large $n$  limit (being $0$ for $r$ less than $n/
(\sqrt{2}\kappa)$ and linked to $g$ otherwise) than for general $n$.
 This is the reason why no new constant needs to be numerically
determined in (\ref{numul1}).

\subsection{The small  $u$ limit $(n\gg \kappa\gg 1)$}

       In the small $u$ case, the length $R$ tends to infinity.
 Defining a local variable $x = r - R$, the differential 
 equation  (\ref{eqlimh})  for $h$ becomes
\begin{equation}
  \frac{ d^2 h}{dx^2} + \frac{1}{R + x} \frac{ d h}{dx}   -\frac{1}{(R+x)^2} h
   = 2 h ( 1 - h^2) ,
\label{eqlimh2}
\end{equation}
 with  the following  boundary conditions 
 \begin{equation}
        h(x=0) = -1  \,\,\,\,\,  \hbox{ and } \,\,\,\,\,
      h(\infty) = 0 .
   \label{bcondh2}
 \end{equation}
 When $R \to \infty$, the equation (\ref{eqlimh2}) reduces to:
 \begin{equation}
  \frac{ d^2 h}{dx^2}    = 2 h ( 1 - h^2) .
\end{equation}
The solution that satisfies the boundary conditions is:
  \begin{equation}
        h(x) = -\sqrt{2}
  \left( 1 - \tanh^2(\sqrt{2}(x + x_0)) \right)^{1/2}
 \,\,\,\,\,  \hbox{ with  } \,\,\,\,\, \tanh(\sqrt{2}x_0) = 
\frac{1}{\sqrt{2}}    . 
 \label{solh}
 \end{equation}
>From this expression, we  derive  that
\begin{equation}
  h'(r=R) = h'(x=0)= 1 .
 \label{hprime0}
\end{equation}
Substituting  this result in (\ref{inverser}) leads to
 the small $u$ behaviour of $R$
 \begin{equation}
    R \simeq \left( \frac{\sqrt{2}}{u} \right)^{1/2}   . 
  \label{Rupetit}
\end{equation}
  The subleading behaviour of $R$ as a function of $u$
 is  found by retaining only the first order term  in $1/R$  in
 (\ref{eqlimh2}):
  \begin{equation}
  \frac{ d^2 h}{dx^2} + \frac{1}{R} \frac{ d h}{dx}
   = 2 h ( 1 - h^2) ,
\label{eqlimh3}
\end{equation}
 with the same boundary conditions as in  (\ref{bcondh2}). This equation can
 be solved perturbatively, but  we  only  need to
  determine   the   correction to equation  (\ref{hprime0}); i.e. we  
  need to calculate the coefficient
 $C$ such that   $ h'(r=R) = 1 + C/R$. We use an  argument
 of energy conservation.
  Defining
 $$ {\cal E}_2 = \frac{1}{2} \left(\frac{dh}{dx}\right)^2 
 - h^2 + \frac{h^4}{2}  , $$
 we  readily  obtain from (\ref{eqlimh3}) that
 $$ \frac{d{\cal E}_2 }{dx} = -\frac{1}{R}\left(\frac{dh}{dx}\right)^2 . $$
Integrating this  relation from $0$ to $\infty$ and
 evaluating the energy $ {\cal E}_2$ at both ends leads to 
\begin{equation}
  C = \int_0^{\infty} \left(\frac{dh}{dx}\right)^2 dx . 
\label{coeffC}
\end{equation}
 Using  the 0th-order  solution (\ref{solh})  for $h$ in (\ref{coeffC}) 
 we obtain 
\begin{equation}
  h'(r=R) = h'(x=0)= 1  + \left( \frac{ 2 \sqrt{2} - 1}{3}\right)
 \frac{1}{R} .
 \label{hprime0bis}
\end{equation}
Substituting  this result in (\ref{inverser}) and solving for $R$ leads to 
 \begin{equation}
    R \simeq \left( \frac{\sqrt{2}}{u} \right)^{1/2}
    \left( 1 -  2^{-1/4}\frac{\sqrt{2} + 1}{3} u^{1/2} +{\cal O}(u)\right) .
  \label{Rupetit2}
\end{equation}
 Substituting the expressions (\ref{Rupetit2})  and (\ref{solh})
  for  $R$  and $h,$  respectively, in (\ref{scalenerg})
  we  obtain  the free energy 
\begin{equation}
\frac{1}{2\pi} {\cal F} \simeq   n \kappa \left( u  R^2 +
  2 R  \int_{0}^{\infty} h^2(x) dx   \right)  \simeq 
     n \kappa \sqrt{2} \left( 1- \frac{4}{3}(\sqrt{2} - 1)2 ^{1/4} 
 \left( \frac{\kappa}{n} \right)^{1/2}
+{\cal O}(u) \right) .
  \label{Enupetit}
\end{equation}
    These results are consistent  with those obtained in section 3. In fact
 the small $u$ limit amounts to first taking the limit
 $n \to \infty$   and then taking $\kappa \to \infty$.
 For instance, it can be shown that equation (\ref{r0ordre1})
 reduces to   (\ref{Rupetit2}) in the $\kappa \to \infty$ limit.
 Moreover,  comparing (\ref{Enupetit}) with  (\ref{approxA}) 
 we  retrieve  the asymptotic behaviour of   the surface tension
 in the large $\kappa$ limit  \cite{sarma,dorsey1}:
\begin{equation}
  \sigma(\kappa) \simeq  - \frac{ 4\sqrt{2}}{3}(\sqrt{2} - 1)\kappa^2 
\simeq -0.7810  \kappa^2  . 
\end{equation}

  \section{A vortex in the small vorticity limit}

  The winding number  $n$ must be a priori
  an integer  because the phase of the wave function $\psi$
  has  to be a $2\pi$  periodic function.
  However, as already noted,
 in the cylindrically  symmetric Ginzburg-Landau equations
  (\ref{gl1},\ref{gl2}),  $n$  appears only  as a parameter in the boundary
  condition at the origin and can be given any real value. In this
  section, we take  $n$ to be  close to 0 and determine
  perturbatively the solutions of (\ref{gl1},\ref{gl2})  and the free energy.

Away from the origin $r=0$, $f$ is close to one and $g$ is
small. Thus, we expand $f$ and $g$ as follows,
\begin{equation}
 f = 1 + n f_1 +n^2f_2 + \ldots \,\,\,\hbox{ and } \,\,\,
  g = n g_1 + n^2g_2 + \ldots
\label{0expansion}
\end{equation}
The functions $f_1$,$g_1$, $f_2$, $g_2$...  satisfy a
 hierarchical system  of linear differential equations
 that can be solved recursively by imposing
 the boundary conditions:  
$ f_i \to 0$ and $  g_i  \to  0  \,\,\,$ when $r \to \infty$, 
for all  $  i \ge 1 .$

 Thus  the first order terms  $f_1$ and $g_1$ are  solutions of
\begin{eqnarray}
    \frac{ d^2 f_1}{dr^2}  + \frac{1}{r}  \frac{df_1}{dr} 
   &=&  4 \kappa^2 f_1 , \label{dgl1}   \\ 
    \frac{ d^2 g_1}{dr^2} -  \frac{1}{r}  \frac{ d g_1}{dr}  &=& 2 g_1 .
  \label{dgl2}
\end{eqnarray} 
 Implementing the boundary and the matching  conditions, we obtain
 \begin{equation}
 f_1 = A K_0(2\kappa r) \,\, , \,\,\,\,  g_1 = B\sqrt{2}r K_1(\sqrt{2} r) ,
\label{petitordre1}
\end{equation}
where $A$ and $B$ are  two undetermined constants  at this stage.

The expansion (\ref{0expansion}) breaks down in the vicinity of the origin
since  $f(0) = 0$. The  expansion of $f$ near $r=0$ 
 is given by (\ref{localf})
\begin{equation}
       f(r)  =  \left(\frac{r}{{\cal R}}\right)^n  + {\cal O}(r^{n+1}) .
 \label{localf2}
\end{equation}
The two expressions (\ref{petitordre1}) and (\ref{localf2}) can be
matched in the parameter range where $r\ll 1$ but $n |\log(r)|\ll 1$.
When $r\ll 1$, the correcting terms in (\ref{localf2}) are negligible
and when  in addition  $n |\log(r)|\ll 1$, one obtains 
\begin{equation}
f = 1 + n \ln\frac{r}{{\cal R}} + {\cal O}(n^2, r) .
 \end{equation}
In the same parameter range, expansion of Eq.~(\ref{petitordre1}) 
gives 
\begin{equation}
f=1 + n A \left( - \ln (\kappa r)  -\gamma + {\cal O}( r^2 \ln r) \right) .
 \label{exlocf}
 \end{equation}
Consistency of these two expressions requires $A=-1$ and
\begin{equation}
{\cal R} = \frac{1}{\kappa}\exp(-\gamma) \simeq  \frac{0.561}{\kappa} ,
\label{formcalR}
\end{equation}
 $\gamma$  being the  Euler constant.

We can proceed in a similar way for the function $g$. Near $r=0$, the
expansion of $g$ is
\begin{equation} 
g = n  \left( - 1 + \left(\frac{r}{r_0} \right)^2 \right)
       - \frac{r^2}{2(n+1)} \left(\frac{r}{{\cal R}} \right)^{2n} 
  +  {\cal O}( r^{2n+4}) \label{devlocalg2} .
\end{equation}
For $r\ll 1$  and 
$n |\log(r)|\ll 1$, this gives
\begin{equation}
g= -n + r^2 \left( -\frac{1}{2}
 + \frac{n}{r_0^2} + \frac{n}{2} -n \ln\frac{r}{{\cal R}}
    + {\cal O}(n^2,r) \right )  . 
\label{exlocg}
\end{equation}
Again, an alternative expression is
obtained from the small $r$ behavior of (\ref{petitordre1})
\begin{equation}
g= n B \left( 1 + r^2 \ln r  + \frac{r^2}{2}
        (  2\gamma  - 1    -\ln 2 )  + {\cal O}( r^2 \ln r)
\right)   .
\label{exbesg}
\end{equation}
Comparing Eqs.~(\ref{exlocg}) and (\ref{exbesg}) gives $B=-1$ and
\begin{equation}
r_0^2 = 2 n - 2 n^2 \ln( 2 \kappa^2) .
\label{coeffmatch}
\end{equation}

Having determined the  small-$n$
 expression for  $f$, the free energy
limit can be calculated  from (\ref{identiteB}). Actually, the expression
(\ref{petitordre1}) for  $f$ is sufficient for this purpose (the range
where it  is not valid gives an exponentially small contribution)  and 
by performing the integral over modified Bessel functions, one obtains
$ {\cal F}/(2 \pi) = n+ O(n^2)$.

 A similar procedure can be carried out for the second order terms.
The expressions for  $f_2$ and $g_2$  are  more intricate
 and   are given in  Appendix B. Using Eq.~ (\ref{identiteB}), they
provide the free energy in the small $n$ limit
up to order $n^2$
\begin{equation}
   \frac{1}{2\pi} {\cal F} = n + n^2 \log(\kappa\sqrt{2}) + {\cal O}(n^3) .
\label{petitfree}
\end{equation}

\section{The dual point of the Ginzburg-Landau Equations}
\subsection{The dual point equations}

  Eq.~(\ref{petitfree}) suggests that the value   $\kappa = 1/\sqrt{2}$ has
 special properties (for instance, for $ \kappa = 1/\sqrt{2}$ 
 the functions  $f_1$ and $g_1$  of   (\ref{petitordre1}) satisfy  
  simple  identities  such as $df_1/dr = -g_1/r$). It is indeed
 well-known  that   at this
 `dual'  point   the second order Ginzburg-Landau equations reduce
 to  first order equations,  leading to 
 special  relations between the functions $f_i$ and $g_i$.
 Moreover, the free energy at the dual point  has been calculated
 exactly and  is identical to the topological number $n$.
 We   recall briefly  these special properties. 

  At  the dual point
  the free energy (\ref{free2}) 
  can be written  as follows:
 \begin{eqnarray}
\frac{1}{2\pi} {\cal F} &=& \int_{0}^{\infty}\left\{
 \left( \frac{df}{dr} +  \frac{fg}{r} \right)^2 +
   \frac{1}{2}  \left( (1 - f^2) - \frac{1}{r} \frac{dg}{dr} \right)^2
\right\} rdr + 
\int_{0}^{\infty} dr \left\{ \frac{d}{dr}( g (1 - f^2) )\right\} 
\nonumber  \\
      &=&    \int_{0}^{\infty}\left\{
 \left( \frac{df}{dr} +  \frac{fg}{r} \right)^2 +
   \frac{1}{2}  \left( (1 - f^2) - \frac{1}{r} \frac{dg}{dr} \right)^2
\right\} rdr  \,\,\, + \,\,\, n  \,\, .
\label{freebogo}
 \end{eqnarray}
 The minimal free energy is obtained when the following first order
 system is satisfied (Bogomol'nyi
 equations \cite{sarma,harden,bogo}) :
  \begin{eqnarray}
 \left( \frac{d}{dr} + \frac{g}{r} \right)f    &=& 0 ,
\label{bogo1} \\
  B = \frac{1}{r} \frac{dg}{dr} &=& (1 - f^2) .
  \label{bogo2} 
\end{eqnarray}
 Substituting (\ref{bogo1}) and (\ref{bogo2}) in (\ref{freebogo})
 we obtain  the  minimal energy 
\begin{equation}
 \frac{1}{2\pi} {\cal F} = n .
 \label{energydual}
 \end{equation}
 Equation (\ref{bogo2}) implies that 
\begin{equation}
  g = -n + \frac{r^2}{2}  - \int_0^r rf^2(r)dr \,\,\, i.e.  \,\,\,
  r_0^2 = 2n  .
\label{r0dual}
\end{equation}
 
 Another remarkable consequence of the Bogomol'nyi
 equations is that the surface energy 
$\sigma(\kappa)$, defined in (\ref{energsurface}), 
 vanishes identically   and changes
 its sign  at the dual point 
 (this is the reason why the dual point separates Type I from
 Type II superconductors \cite{sarma}):
\begin{equation}
 \sigma(\kappa = \frac{1}{\sqrt{2}}) = 0 .
\end{equation}

 All the calculations that we have carried out in the previous
 sections are consistent with these (non-perturbative) properties
 of the dual point. In the small $n$ case of section 5, one can
 verify,  using (\ref{petitordre1}) and 
 the expressions given in the Appendix,  
 that the equations (\ref{bogo1}, \ref{bogo2}) are
 satisfied order by order by the 
 expansions (\ref{0expansion})   of $f$ and $g$ 
 at the dual point. Moreover,   the $n^2$
 correction to the free energy in (\ref{petitfree}) 
 and to $r_0$ in  (\ref{coeffmatch})
 vanishes  identically    at $\kappa = {1}/{\sqrt{2}}$,
 as expected. In the large $n$ case,  the  expression (\ref{approxA}) 
 for the free energy reduces to  (\ref{energydual}) because 
 the $n^{1/2}$ correction disappears thanks to the  vanishing  of 
 the surface energy;  for the same reason,   
 the formula (\ref{r0ordre1}) simplifies  to  $r_0 = (2n)^{1/2}$.

 At the dual point,   the   vortices
 do not interact and  the free energy does not depend
 upon their location. This makes it possible to obtain
 exact $n$-vortices solutions, in the large $n$
 limit,  for arbitrary locations of the vortex
 cores \cite{efanov}. 
 However, when  
 $\kappa$ deviates from the special  value
 $1/\sqrt{2}$, vortices  start  interacting and   their 
 interaction energy is  extremal when all the vortices
 are located at the same point. We now evaluate
 the energy of such a configuration with cylindrical symmetry
 when $\kappa$ is close to the dual point.

 \subsection{Free energy of a giant vortex  in the vicinity of  the dual point}

       We apply our  results 
  to the case where  $\kappa$  is  close to the   dual point
 value $\kappa = 1/{\sqrt 2}.$   
  In experiments on   mesoscopic superconductors,  values of 
 $\kappa$  are close to the dual point  value  \cite{geim1,akmal2}
 and in previous theoretical studies the special case $\kappa = 1/{\sqrt 2}$
 was found to be  very useful  to study analytically 
 the magnetization of a sample as a function of  the applied field.
 In the vicinity of the dual point, a   relevant  quantity 
  is  the  relative  growth
  of the  free energy:
\begin{equation}
 \alpha(n)  =\frac{1}{\kappa\sqrt{2} - 1}
  \frac{{\cal F}(\kappa, n) - {\cal F}(1/\sqrt{2} , n) }
{{\cal F}(1/\sqrt{2} , n) }  \,\,\,  \hbox{ when }\,\,\,
  \kappa   \to  \frac{1}{\sqrt{2}}         .
\label{defalpha}
\end{equation} 
 This relation  is a local form  of an empirical
  scaling of the free energy
 \begin{equation}
\frac{1}{2\pi} {\cal F}(\kappa, n) \simeq  n (\kappa\sqrt{2})^{\alpha(n)} ,
 \label{exposants}
 \end{equation} 
 found  in  \cite{akmal2}, where 
  the exponents $\alpha(n)$ were  computed numerically.
 Differentiating   the Ginzburg-Landau free energy (\ref{free2}) with respect
 to the parameter $\kappa$
    at the dual point  we obtain:
   \begin{equation}  
      \alpha(n)  = \frac{1}{n} \int_{0}^{\infty} 
    (1 - f^2)^2 rdr   =   \frac{1}{n} \int_{0}^{\infty} 
  \frac{dr}{r} \left( \frac{ dg}{dr}\right)^2            \,\,\,    . 
\label{defalpha2}
\end{equation}
 The last relation was derived from  the Bogomol'nyi equation (\ref{bogo2}).

  We now explain how a large $n$ expansion of the  exponents $\alpha(n)$ 
 can be derived from our previous results. Integrating Eq.~ (\ref{defalpha2})
 by parts,  we find
\begin{equation} 
 \alpha(n)  = \frac{1}{n}  \int_{0}^{\infty} \frac{ r^2}{2} \psi(r) dr 
  \,\,\,  \hbox{ with } \,\,\,  \psi =  4 f(1 - f^2) \frac{df}{dr}  .
\label{identitealp3} 
\end{equation}
 The function $\psi(r)$ is localized at  the position of the front
 and its width is of the order $1 /\kappa$. Hence   we can 
  expand (\ref{identitealp3})  using the local variable
 $ x = r -r_0$  and  in the large $n$ limit  replace
  $f$  by $f_0$ (as in section 3.5):
 \begin{equation}
  \alpha(n)  = \frac{1}{n} \frac{r_0^2}{2} + \frac{r_0}{n} 
 \int_{-\infty}^{\infty} x  \psi(x) dx \,\,\,\, \hbox{ with } \,\,\,
  \psi =  4 f_0(1 - f_0^2)\frac{df_0}{dx}  .
\label{devptalpha}
\end{equation}
 Integrating the last term by parts, we obtain:
 \begin{eqnarray}
\int_{-\infty}^{\infty} x  \psi(x) dx &=& 
 \int_{-\infty}^0  dx \left\{ (1 - f_0^2)^2
 - 1 \right\} +  \int_{0}^{\infty} dx  (1 - f_0^2)^2 \nonumber \\
 &=& \int_{-\infty}^{\infty}  dx \left\{(1 - f_0^2)^2 - \frac{dG_0}{dx} 
  \right\}
 =  - \int_{-\infty}^{\infty} f_0^2(1 - f_0^2) dx .
 \label{ippsi}
\end{eqnarray}
 In  the last equality we used the fact that the functions $f_0$
 and $G_0$ satisfy the following  Bogomol'nyi equations
 at the dual point \cite{dorsey2}:
\begin{equation}
  \frac{dG_0}{dx}  = (1 - f_0^2) \,\,\,\, \hbox{ and } \,\,\,
\frac{df_0}{dx} + G_0 f_0   = 0 . 
\end{equation}
 The function $v$ defined by $f_0 = \exp( - v/2) , $  satisfies a second order
 differential equation that can be  solved  explicitely \cite{dorsey2}:
 \begin{equation}
 \frac{1}{2}\frac{d^2 v }{dx^2} = 1 -   \exp( - v )    
 \,\,\,\, \hbox{ which implies  } \,\,\, 
 \frac{dv}{dx}  =  -2 \left(  e^{ - v }   - 1 + v \right)^{1/2}  .
\end{equation}
Now changing the variable $x$  to  $v$,
we rewrite the last integral in  (\ref{ippsi}) as
 \begin{equation}
 - \int_0^\infty dv  \frac{ e^{-v} ( 1 - e^{-v}) dv}
 { (  e^{ - v }   - 1 + v )^{1/2} } = - 
 \int_0^\infty dv e^{-v} \left( e^{-v} - 1 + v\right)^{1/2} 
 \simeq  -  0.5482  . 
 \label{calculint}
\end{equation}
  Substituting  this result in (\ref{devptalpha})
 and knowing  that  $r_0^2 = 2n$ exactly (\ref{r0dual}),
 we find  an asymptotic expansion for
  $\alpha(n)$  in the large $n$ limit:
\begin{equation}
\alpha(n) =  1 - \frac{0.7753}{n^{1/2}} + {\cal O} \left(\frac{1}{n} \right) .
\label{ordre1}
\end{equation} 
We have not calculated  the subleading behaviour of the free energy but
our numerical results  provide  an estimate of
the higher  order  term in (\ref{ordre1}):
\begin{equation}
\alpha(n) =  1 - \frac{0.7753}{n^{1/2}} +  \frac{{\cal K}}{n}
 + {\cal O} \left(\frac{1}{n^{3/2}} \right)  \,\,\, \hbox{ with }
 \,\,\, {\cal K}\simeq  0.125 .
\label{ordre2}
\end{equation}
 From (\ref{ordre1})  and (\ref{defalpha}),  
 the   expansion  of the free energy near  $\kappa = \frac{1}{\sqrt{2}}$
 is found. This allows us to retrieve, 
  using (\ref{approxA}),   the local behaviour   of
  the surface energy in the vicinity of the dual point,
  first calculated  by Dorsey \cite{dorsey2}:
\begin{equation}
 \sigma(\kappa)  \simeq    -  (\kappa\sqrt{2} - 1)\int_0^\infty dv
 \left( e^{-v} - 1 + v\right)^{1/2}e^{-v} \simeq  -0.5482(\kappa\sqrt{2} - 1) .
\end{equation}

In the same way,  from the small $n$ expansion of the free energy 
up to the second order,  we derive that  (see Appendix B): 
\begin{equation}
\alpha(n) = n  - 4 n^2\int_{0}^{\infty} r^2 K_0^2(r)K_1(r)dr +  {\cal O }(n^3)
    \simeq n - 1.5626 n^2 + {\cal O }(n^3) . 
\label{devordre2petit}
   \end{equation}
 We have computed  numerically  from (\ref{defalpha})
  the function  $\alpha(n)$ for
 $n$ ranging from 1 to 250   and  some
 significant  values are given in Table 1.
 The numerically computed and 
  the  large $n$ asymptotic expansion
 (\ref{ordre1}) of $\alpha(n)$
 agree to  better than 7$\%$ 
 even for $n$ as small as 4.   In particular we notice from  (\ref{ordre1})
 that $\alpha(n) \to 1$ when $n \to \infty$. In the $ n \to 0$ limit,
 the second order expansion (\ref{devordre2petit}) provides
 a fairly good approximation for $n \le 0.25$. In Fig. 7, 
 numerically computed values of $\alpha$ are compared to large $n$ and
 small $n$ expansions.

\begin{table}
  \centering 
  \begin{tabular}{||c|c||c|c||c|c|} 
  \hline 
 &                &         &                &      &    \\ 
 n   &   $\alpha (n)$ &     n   &  $\alpha(n)$   &  n   &  $\alpha(n)$\\ 
     &                &         &                &      &    \\ 
\hline 
1  & 0.415    & 12  & 0.789    & 50  & 0.893    \\
2  & 0.542    & 13  & 0.797    & 60  & 0.902    \\
3  & 0.611    & 14  & 0.804    & 80  & 0.915    \\      
4  & 0.655    & 15  & 0.810    & 100 & 0.924    \\          
5  & 0.687    & 16  & 0.816    & 120 & 0.930    \\  
6  & 0.711    & 17  & 0.821    & 140 & 0.935    \\
7  & 0.730    & 18  & 0.826    & 160 & 0.940    \\
8  & 0.746    & 19  & 0.830    & 180 & 0.943    \\
9  & 0.759    & 20  & 0.834    & 200 & 0.946    \\
10 & 0.770    & 30  & 0.863    & 225 & 0.949   \\ 
11 & 0.780    & 40  & 0.881    & 250 & 0.951    \\                           
\hline 
\end{tabular}
\hspace{10pt} 
\caption{Numerical values of the function $\alpha (n)$ 
  obtained from (\ref{defalpha2}) for  different  $n$ 
 in the range 1 to 250}   
\end{table}

\begin{figure}
\begin{center}
{\psfig{figure= 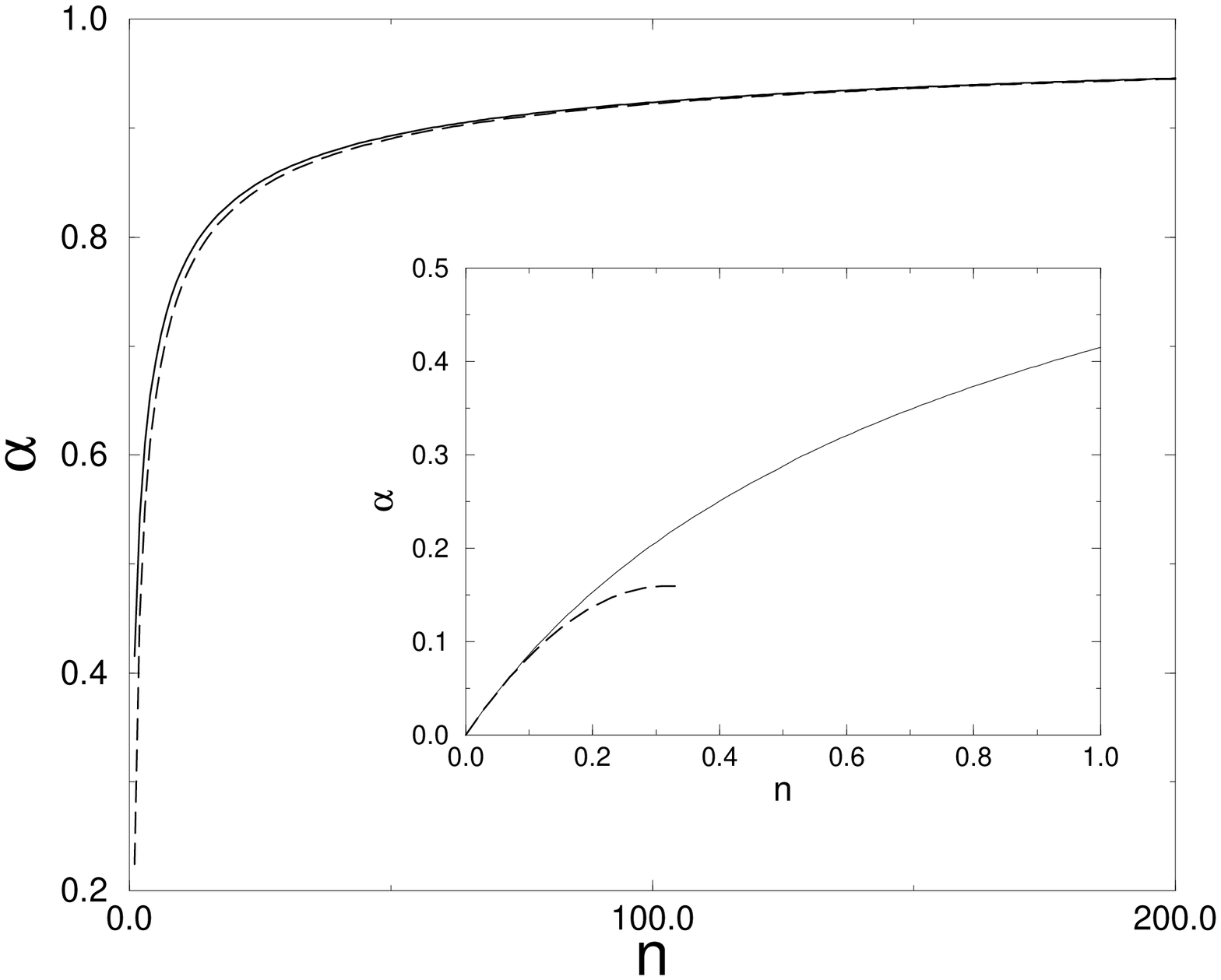,height=10cm,angle=-90}}
\end{center} 
\caption{The function $\alpha(n)$ is plotted for $n$ between 1 and 200,
 and  in inset for values of $n$ between 0 and 1 (solid line curves).
  The dashed lines represent the asymptotic expansions for 
  large $n$ (\ref{ordre1})  and small $n$  (\ref{devordre2petit}). }
\label{}
\end{figure}

\section{Conclusion}

We have analyzed 
 the structure of a  giant vortex of winding number
$n$ in an infinite plane
for arbitrary  values of the parameter $\kappa$  in contrast to   previous
analytical  results  obtained only  in  the London limit or at the dual point. 
 The vortex and magnetic field profiles
  are  computed   by solving the Ginzburg-Landau equations  which
 minimize the free energy of the system. These numerical solutions are 
compared  with  analytical  results derived  in
the cases where  the vortex
multiplicity $n$ is  either very
large or very small. In particular,
a simple structure has been found for large $n$, its relation to the classic
result of Abrikosov  has been
 elucidated and the perturbative expansions  are  found to agree
well with previous numerical computations of a giant vortex free energy. 
We hope that some of these results will prove useful for the current 
very active
experimental investigations of 
mesoscopic superconductors.

{\bf Acknowlegments.} This work was initiated when two of us (V.H. and K.M.)
were attending
the workshop "Ginzburg-Landau models, vortices and low-temperature physics"
at the Lorentz Institut, Leiden. It is a pleasure to thank the organizers,
 of this conference,  particularly Eric Akkermans, for stimulating discussions
 and the Lorentz Institut for its hospitality. K.M. would like to thank
 S. Mallick for his constant help during the preparation of this work.

\appendix

\section{expressions of the normal/superconducting interface energy}
\label{stapp}
The energy  per  unit area of a planar normal/superconducting interface
is obtained from the Ginzburg-Landau energy (\ref{energ}) under the form
\cite{ll}
\begin{equation}
\sigma(\kappa)=\int_{-\infty}^{\infty}\left\{\left(\frac{df_0}{dx}\right)^2
+
 f_0^2G_0^2 +
  \kappa^2 (1 - f_0^2)^2 + \frac{1}{2}\left(\frac{dG_0}{dx}\right)^2 
-\kappa \sqrt{2}\frac{dG_0}{dx} \right\} dx ,
\label{st1}
\end{equation}
where $f_0$ and $G_0$  satisfy 
 Eqs.~(\ref{inngl1}, \ref{inngl2}). We notice that
for a  magnetic field oriented along the $z$-direction, $G_0$  represents
the $y$-component of the potential vector.
The expression (\ref{st1}) can be  written  in different  forms by
making use of the identities:
\begin{eqnarray}
\int_{-\infty}^{\infty}\left(\frac{df_0}{dx}\right)^2 dx&=&
\int_{-\infty}^{\infty}\left\{2\kappa^2 f_0^2
   ( 1 - f_0^2)-G_0^2 f_0^2
  \right\} dx , 
\label{id1}\\
\int_{-\infty}^{\infty}dx \left\{\left(\frac{dG_0}{dx}\right)^2
-\kappa \sqrt{2}\frac{dG_0}{dx}\right\}&=&-2
 \int_{-\infty}^{\infty}G_0^2 f_0^2
 dx , 
\label{id2} \\
 G_0^2 f_0^2 &=&  \left(\frac{df_0}{dx}\right)^2 +
  \frac{1}{2}\left(\frac{dG_0}{dx}\right)^2 - \kappa^2 (1 - f_0^2)^2  . 
\label{id3}
\end{eqnarray}
The first two   identities are obtained (e.g. \cite{dorsey1})
 from  Eqs.~(\ref{inngl1},\ref{inngl2})   by integration 
by parts. 
The third identity results  from   the conservation law (\ref{hamiltonien}).
 Using (\ref{id1}) to eliminate
$\int (df_0/dx)^2$ from Eq.~(\ref{st1})  provides 
the expression (\ref{energsurface}) of the main text. Similarly,   using 
Eqs.(\ref{id1},\ref{id2},\ref{id3}) to eliminate $\int
(df_0/dx)^2,  \int (dG_0/dx)^2$ and 
 $\int f_0^2 G_0^2\,$   from Eq.~(\ref{st1})   leads to 
 the alternative expression (\ref{formulesimple})  of
 the main text:
\begin{equation}
 \sigma(\kappa) =  \frac{ 4}{3}\kappa^2
 \int_{-\infty}^{\infty} \left( (1 - f_0^2) - \frac{1}{\kappa\sqrt{2}}
 \frac{d G_0} {d x}  \right) dx   .
 \label{altexpr}
 \end{equation}
Finally, we note that the integral of $x\,\phi$ in
(\ref{formulesimple})
 is transformed
into Eq. (\ref{altexpr}) 
  by integrating separately by parts the
integrals between $-\infty$ and $0$ and between $0$ and $+\infty$:
 \begin{eqnarray} 
   \int_{-\infty}^{\infty}\phi(x) x dx =   
 \int_{-\infty}^{\infty} \left( (x+x_c)  \frac{d f_0^2}{dx}  \right) dx  
=   \int_{-x_c}^{\infty}\left( 1 - f_0^2  \right) dx  - 
  \int_{-\infty}^{-x_c}  f_0^2   dx  \nonumber\\
 = 
 \int_{-\infty}^{\infty} \left( (1 - f_0^2) - \frac{1}{\kappa\sqrt{2}}
 \frac{d G_0} {d x}  \right) dx  
   \end{eqnarray} 
(the last equality is obtained by adding $0$ under the complicated form
\begin{equation}
0=
\int_{-x_c}^{+\infty}dx[-\frac{1}{\kappa\sqrt{2}}
\frac{dG_0}{dx}]+\int_{-\infty}^{-x_c}
dx [1-\frac{1}{\kappa\sqrt{2}}\frac{dG_0}{dx}]\ ).
\nonumber
\end{equation}

\section{Second order computation in the $n->0$ limit and some useful formulas}

  We first recall  some useful asymptotic
 formulae for the modified Bessel functions:
  \begin{eqnarray} 
   K_0(r)  &=&  - \ln \frac{r}{2} -\gamma +  {\cal O}( r^2 \ln r) ,
  \nonumber \\
   K_1(r)   &=&   \frac{1}{r}  +\frac{r}{2} \ln\frac{r}{2}  +\frac{r}{4}
   (  2\gamma  - 1)  +  {\cal O}( r^2 \ln r)  ,  \nonumber \\
    K_2(r)  &=&  \frac{2}{r^2} +  {\cal O}( r^{-1})  . 
 \label{formBessel}
\end{eqnarray} 
 \begin{eqnarray} 
 \frac{d K_0}{dr}   =   - K_1(r) \,\,\, &\hbox{and}& \,\,\,
    \frac{d}{dr}( r K_1(r)  )  =   - r K_0(r)   .  \nonumber \\
 \frac{d I_0}{dr}   =   I_1(r)
 \,\,\,  &\hbox{and}& \,\,\,  \frac{d}{dr}( r I_1(r)  ) =  r I_0(r)  . 
  \label{derivBessel}
 \end{eqnarray}
 We now  explain how the matching procedure
 is carried out up to the second order 
 in the case  $n \to 0.$
 The functions  $f_2$ and $g_2$   
 solve   the linear system: 
  \begin{eqnarray}
 \frac{ d^2 f_2}{dr^2}  + \frac{1}{r}  \frac{df_2}{dr} 
  &=& 4 \kappa^2 f_2 +   \frac{g_1^2}{r^2} + 6\kappa^2 f_1^2 ,
 \nonumber  \\ 
\frac{ d^2 g_2}{dr^2} -  \frac{1}{r} \frac{d g_2}{dr}  &=& 2 g_2 + 4f_1g_1 .
  \nonumber  
\end{eqnarray} 
 Solving this system by   `variation of constants', 
 we obtain:
\begin{eqnarray}
 f_2 =  A  K_0(2\kappa r) + B I_0(2\kappa r)   
   &-& I_0(2\kappa r)  \int_{ r}^{\infty}  
  u du K_0(2\kappa u) \left( 2  K_1(u\sqrt{2})^2  + 6 \kappa^2
   K_0(2\kappa u)^2 \right)   \nonumber   \\ 
    &-& K_0(2\kappa r)
  \int_{ 1}^{r}   u du I_0(2\kappa u)\left( 2  K_1(u\sqrt{2})^2  + 6 \kappa^2
   K_0(2\kappa u)^2 \right) , \label{f2general}  
   \end{eqnarray} 
\begin{eqnarray}
   g_2 =  C  K_0(2\kappa r) + D  I_0(2\kappa r)    
    && -4\sqrt{2}r  I_1(\sqrt{2} r)
   \int_r^{\infty}K_0(2\kappa u)(K_1(\sqrt{2} u))^2 udu   \nonumber   \\   
  &&  -4\sqrt{2}r K_1(\sqrt{2} r) \int_0^{r} K_0(2\kappa u) K_1(\sqrt{2} u)
  I_1(\sqrt{2} u)u du  \label{g2general}   . 
\end{eqnarray} 
 Since $f_2$ and $g_2$ must be finite at infinity, we have 
 $B = D = 0$. The two other coefficients are  found  by
 matching  with the inner expansions in the vicinity of zero.
 For this purpose, we need  the local expansions
 of $f$ and $g$ in the vicinity of 0 up to the order $n^2$.
 Using (\ref{localf2}) and (\ref{devlocalg2}), we obtain
 \begin{eqnarray}
  f(r)  &=&  1 + n \left(  \ln\frac{r}{{\cal R}} + {\cal O}(r \ln r)\right)
  + \frac{n^2}{2} \left(  (\ln\frac{r}{{\cal R}})^2 +  {\cal O}(r \ln r)\right)
   +  {\cal O}(n^3)  .  \label{locfappB} \\
  g(r) &=& -n \left( 1 + {\cal O}(r ^2\ln r) \right)   
   +  n^2 \left(  r^2\ln\frac{r}{{\cal R}}(1 - \ln\frac{r}{{\cal R}}) 
 + {\cal O}(r ^2) \right)  +  {\cal O}(n^3)   . \label{locgappB} 
\end{eqnarray}
 Because  the coefficient of $n^2$ in  the local 
 expansion  (\ref{locgappB})  of $g$  tends  to zero
 when  $r$ vanishes, one must have  $ g_2 \to 0 $  when $ r \to 0$
 and therefore  $C = 0$. 
 To obtain  the coefficient $A$, we must determine  the divergences of 
 $f_2$  given by (\ref{f2general})  when  $ r \to 0$:
 the term  $ A  K_0(2\kappa r)$   produces a singular term
  $- A \ln r  $ when  $ r \to 0$; 
  from the  expansions of the Bessel 
 functions (\ref{formBessel}), we  find   the diverging part of
  the first integral  in (\ref{f2general})  to be    
$  - \int_{ r}^{\infty}  
 2 u du K_0(2\kappa u)  K_1(u\sqrt{2})^2 $
 which is equivalent  to
  $   \int_{ 2\kappa r}^{1} {(\ln \frac{u}{2} + \gamma) }du/{u} ; $
 thus   the first integral  gives  a singular term equal to 
$$ - \frac{(\ln r)^2}{2} + (-\gamma -\ln\kappa) \ln r  \, ; $$
 similarly the  singular term  due to the second integral 
 in (\ref{f2general})  is   given by:
$$   (\ln r)^2 - \ln r \left( -\gamma -\ln\kappa 
 + 2 \int_0^1  u du \left( I_0(2\kappa u) K_1(u\sqrt{2})^2 - 
 \frac{1}{2 u^2}\right)\right). $$
 Hence the total  diverging part of $f_2$ is:
  \begin{equation} 
    \frac{(\ln r)^2}{2}  - \ln r  \left( A + 2 \int_0^1 
 u du \left( I_0(2\kappa u)  K_1(u\sqrt{2})^2 - \frac{1}{2 u^2}\right)\right) .
  \end{equation}
 Identifying this expression with  the singular part of   the $n^2$
   term in the expansion (\ref{locfappB})  of $f$
   and using (\ref{formcalR})  provides the value of $A$
   and  leads to   the second order
 terms in the small $n$ expansion 
 of  $f$ and $g$:
\begin{eqnarray}
 f_2 =  A  K_0(2\kappa r)  - I_0(2\kappa r)  \int_{r}^{\infty}  
  u du K_0(2\kappa u) \left( 2  K_1(u\sqrt{2})^2  + 6 \kappa^2
   K_0(2\kappa u)^2 \right)   &&  \nonumber   \\ 
    - K_0(2\kappa r)
  \int_{ 1}^{r}   u du I_0(2\kappa u)\left( 2  K_1(u\sqrt{2})^2  + 6 \kappa^2
   K_0(2\kappa u)^2 \right) ,   
 &&  \nonumber \\
 \hbox{ with  } \,\,\,\, A   =  -\gamma  - \ln\kappa - 2 \int_0^1 
 u du \left( I_0(2\kappa u)  K_1(u\sqrt{2})^2 - \frac{1}{2 u^2}\right)  
   && \label{formulef2eme}  ; \end{eqnarray}  
 \begin{eqnarray}
  g_2  =   &-& 4\sqrt{2}r  I_1(\sqrt{2} r)
   \int_r^{\infty}K_0(2\kappa u)(K_1(\sqrt{2} u))^2 udu   \nonumber   \\  
    &-&  4\sqrt{2}r  K_1(\sqrt{2} r) \int_0^{r} K_0(2\kappa u)K_1(\sqrt{2} u)
  I_1(\sqrt{2} u)u du   .
 \label{formuleg2eme}
\end{eqnarray}

 We can now derive the small $n$ asymptotic behaviour of 
 $\alpha(n)$. From the explicit relation (\ref{defalpha2}), we obtain,
 up to the second order in $n$:
\begin{equation}  
      \alpha(n)  =  \frac{1}{n} \int_{0}^{\infty} 
  \frac{dr}{r} \left( \frac{ dg}{dr}\right)^2  
  =   n  \int_{0}^{\infty}  \frac{dr}{r} \left( \frac{ dg_1}{dr}\right)^2
 + 2 n^2  \int_{0}^{\infty}  \frac{dr}{r}
  \left( \frac{ dg_1}{dr}\frac{ dg_2}{dr}\right)  + {\cal O}(n^3)  . 
 \label{localpha2}
\end{equation}
 where the functions $g_i$ are calculated at the dual point 
 $\kappa = 1/\sqrt{2}$.
 Recalling  that  $g_1 = - \sqrt{2} r K_1(\sqrt{2} r)$ 
  (see \ref{petitordre1})  and using (\ref{derivBessel}) we have 
 \begin{equation}  
\int_{0}^{\infty} \frac{dr}{r} \left( \frac{dg_1}{dr}\right)^2 = 
  2 \int_{0}^{\infty} x K_0^2(x) dx = 1  .  
\end{equation}
 From  (\ref{derivBessel}) 
 and  (\ref{formuleg2eme}),   we obtain ${ dg_1}/{dr} = 2 r  K_0(\sqrt{2} r)$
 and 
 \begin{equation} 
 \frac{ dg_2}{dr} = 4r \left\{- I_0(\sqrt{2} r)
   \int_{\sqrt{2}r}^{\infty}K_0(u)(K_1(u))^2 udu 
  +   K_0(\sqrt{2} r) \int_0^{\sqrt{2}r} K_0(u)K_1(u)
  I_1(u)u du   \right\} . 
\end{equation} 
 The coefficient of $n^2$ in  (\ref{localpha2}) is therefore given by:
  \begin{eqnarray}
  2 \int_{0}^{\infty}  \frac{dr}{r}
  \left( \frac{ dg_1}{dr}\frac{ dg_2}{dr}\right) =  
 4 \int_{0}^{\infty}  dr  K_0(\sqrt{2} r)\frac{ dg_2}{dr} 
 = &-& 8\int_0^{\infty} dv \,  v  K_0(v)I_0(v) 
 \int_{v}^{\infty}K_0(u)(K_1(u))^2 udu  \nonumber \\
 &+&  8 \int_0^{\infty} dv \,   v  K_0^2(v)
\int_0^{v} K_0(u)K_1(u)I_1(u)u du   .   \label{interform}
\end{eqnarray} 
 From (\ref{derivBessel})  we can verify that 
  \begin{equation} 
 \int dv  \, v  K_0^2(v) = \frac{v^2}{2} (  K_0^2 -  K_1^2 ) \,\,\, 
    \hbox{ and } \,\,\,   \int  dv \,   v  K_0(v)I_0(v) 
 = \frac{v^2}{2}(  K_0 I_0 +   K_1  I_1 ) . 
 \end{equation} 
Integrating (\ref{interform}) by parts  leads to:
 \begin{eqnarray}
-  8 \int_0^{\infty} dv \,   v  K_0(v)I_0(v) 
 \int_{v}^{\infty}K_0(u)(K_1(u))^2 udu &=&  -  4  \int_0^{\infty}
  dv  \,  v^3(  K_0 I_0 +   K_1  I_1 ) K_0 K_1^2  ,  \nonumber \\
  8  \int_0^{\infty} dv  \,  v  K_0^2(v)
\int_0^{v} K_0(u)K_1(u)I_1(u)u du  &=&  -4  \int_0^{\infty}
 dv  \,  v^3( K_0^2 -  K_1^2 ) K_0 K_1 I_1    . 
 \end{eqnarray}  
Summing these two expressions  and using  $K_0I_1 + K_1I_0 = 1/v$
leads to 
 \begin{equation} 
  2 \int_{0}^{\infty}  \frac{dr}{r}
  \left( \frac{ dg_1}{dr}\frac{ dg_2}{dr}\right) =
   -  4 \int_{0}^{\infty}  dv \,  v^2 K_0^2(v)K_1(v)   . 
\end{equation}
 This concludes the derivation of (\ref{devordre2petit}).
  The free energy in the small $n$ limit
  up to order $n^2$ (\ref{petitfree})  is derived
  by  similar calculations.


\begin{thebibliography}{article}
\bibitem{note1} Relation (\ref{localf}) shows that the order parameter vanishes
 $n$ times at the origin, i.e. that there is indeed  a  giant vortex
 of order $n$ at the origin.
\end{thebibliography}
\end{document}